\documentclass[apj]{emulateapj}
\pdfoutput=1

\begin{document}
\title{Physical Properties of Sub-galactic Clumps at 0.5 $\leq z \leq$ 1.5 in the UVUDF}

\author{Emmaris Soto\altaffilmark{1} and Duilia F. de Mello\altaffilmark{1}}
\affil{Physics Department \\
The Catholic University of America \\
620 Michigan Ave. NE \\
Washington, DC 20064, USA}

\author{Marc Rafelski\altaffilmark{1}}
\affil{Space Telescope Science Institute \\
3700 San Martin Drive \\
Baltimore, MD 21218, USA}

\author{Jonathan P. Gardner}
\affil{NASA Goddard Space Flight Center \\
Observational Cosmology Laboratory \\
Greenbelt, MD 20771, USA}

\author{Harry I. Teplitz}
\affil{Infrared Processing and Analysis Center, Caltech \\
Pasadena, CA 91125, USA}

\author{Anton M. Koekemoer, Swara Ravindranath, and Norman A. Grogin}
\affil{Space Telescope Science Institute \\
3700 San Martin Drive \\
Baltimore, MD 21218, USA}

\author{Claudia Scarlata}
\affil{Minnesota Institute of Astrophysics \& School of Physics and Astronomy \\
University of Minnesota \\
Minneapolis, MN 55455, USA}

\and

\author{Peter Kurczynski and Eric Gawiser}
\affil{Department of Physics and Astronomy \\
Rutgers University \\
Piscataway, NJ 08854, USA}

\altaffiltext{1}{NASA Goddard Space Flight Center}

\begin{abstract} 
We present an investigation of clumpy galaxies in the Hubble Ultra Deep Field at 0.5 $\leq z \leq$ 1.5 in the rest-frame far-ultraviolet (FUV) using HST WFC3 broadband imaging in F225W, F275W, and F336W.  An analysis of 1,404 galaxies yields 209 galaxies that host 403 kpc-scale clumps.  These host galaxies appear to be typical star-forming galaxies, with an average of 2 clumps per galaxy and reaching a maximum of 8 clumps.  We measure the photometry of the clumps, and determine the mass, age, and star formation rates (SFR) utilizing the SED-fitting code FAST.  We find that clumps make an average contribution of 19\% to the total rest-frame FUV flux of their host galaxy.  Individually, clumps contribute a median of 5\% to the host galaxy SFR and an average of $\sim$4\% to the host galaxy mass, with total clump contributions to the host galaxy stellar mass ranging widely from less than 1\% up to 93\%.  Clumps in the outskirts of galaxies are typically younger, with higher star formation rates, than clumps in the inner regions.  The results are consistent with clump migration theories in which clumps form through violent gravitational instabilities in gas-rich turbulent disks, eventually migrate toward the center of the galaxies, and coalesce into the bulge.
\end{abstract}

\keywords{galaxies: evolution, formation, star formation, structure}

\section{Introduction} \label{sec:intro}

The build up of stars in galaxies along the Hubble sequence and subsequently the evolution of those galaxies as observed today remains uncertain in extragalactic astronomy.  Past studies, targeting primarily high redshift galaxies, found increasingly irregular, asymmetric, and clumpy structures in star-forming galaxies \citep{Im1999, Driver1995, Driver1998, Bergh1996, Abraham1996, Glazebrook1995}.  ``Normal'' star-forming galaxies were in place at $z \sim 0.5$, with stellar populations and scaling relations consistent with gradual evolution into the population of galaxies observed locally (e.g. \citealt{Sargent2007, Scarlata2007}).  Looking back to $z > 2$, dramatic changes appear.  Studies have shown that galaxies at these high redshifts are dominated by irregular and peculiar galaxies \citep{Conselice2005, Abraham1996} that have no obvious similarity in terms of structure to lower redshift galaxies \citep{Cameron2010, Overzier2010, Cassata2005, Lotz2004}.  They become more clumpy at increasing redshift \citep{Elmegreen2004} as a result of mergers and other processes which lead to violent gravitational instabilities \citep{Wisnioski2011, Ceverino2010, Genzel2008, Bournaud2007}.  Massive galaxies along the so-called star-forming main sequence \citep[MS;][]{Noeske2007} at these epochs tend to be thick, clumpy disks, forming stars at rates (100 M$_{\odot}$/yr) much higher than is observed in the thin, quiescent, Milky-Way-like disks at $z < 0.5$ (e.g. \citealt{Genzel2008}).  To map this important transition, it is crucial to follow the star formation history (SFH) of individual substructures at intermediate redshifts ($0.5\leq z\leq 1.5$).

High resolution imaging has shown that kpc-sized clumps appear to be a common feature of galaxies at intermediate redshifts, and simulations indicate that they form in-situ by gravitational instabilities in gas-rich galaxies \citep{Noguchi1999, Elmegreen2004, EE2005, Conselice2004, Papovich2005, Bournaud2007, Agertz2009, Ceverino2010}.  Clumps at $z \sim 2$ can reach 10\textsuperscript{9} M$_{\odot}$ \citep{Guo2012, Tacconi2013}; however, their eventual fate remains uncertain.  If they are long-lived (with lifetimes comparable to the orbital timescale of the disk), clumps can migrate inward and provide a path towards bulge growth \citep{Ceverino2010, Bournaud2007}.  It is also possible that powerful outflows could disrupt clumps on short timescales, implying that secular bulge growth would occur more slowly \citep{Genel2012-jan, Forster2011-sep2}.

Clumps are mostly identified in optical imaging from the Hubble Space Telescope (HST) at $z\sim2$ probing the near-UV (NUV).  However, the rest-frame FUV 1500{\AA} is a vital tracer of star formation, directly sampling light from young hot stars \citep{Calzetti2013}, and is thus the best way to identify star-forming clumps.  Therefore, FUV studies are essential for the study of the formation and evolution of galaxies.

Rest-frame FUV data of star-forming clumps at intermediate redshifts has not been explored sufficiently well, making the high resolution UV imaging of the HST Wide Field Camera 3 (WFC3) of the Hubble Ultra Deep Field (HUDF; \citealt{Teplitz2013}) a unique data set for our study.  This epoch is crucial to test the late stage evolution of clumps and disks against competing models.  At 0.5 $\leq z \leq$ 1.5, massive star-forming galaxies are rare due to the exponential drop of the stellar mass function, and thus few are available \citep{Drory2008}.  Massive galaxies with giant UV clumps are even fewer (15-20\%) at $z \sim 1$ \citep{Guo2015}.  However, since clumps have high UV luminosity to stellar mass ratios they are prominent in UV images.

In this paper, we identify clumps, measure their UV sizes, determine the total number of clumps per galaxy, rest-frame 1500{\AA} flux, and constrain stellar mass and stellar population properties.  From these properties we explore the potential fates of the clumps described by two different scenarios: (1) the inward migration and bulge growth scenario and (2) the quick disruption scenario.  In the first case, clumps in disks migrate toward the center of the potential well of the galaxy and coalesce to form a bulge.  Therefore, clumps closer to the center of the galaxy in this scenario are older and denser than those in galaxy outskirts \citep{Bournaud2014}.  In the second case, if feedback is strong, young clumps that exist over the entire galaxy could dissipate and form the disk \citep{Bournaud2008, Oklopcic2016}.  The measured physical properties of the clumps allow us to infer which of these dominate the clump population presented in this study.

The paper is organized as follows: in Section~\ref{sec:data} we summarize the observations that comprise our data set.  In Section~\ref{sec:clumps}, we provide our clump definition and the four criteria which detections must comply with in order to be designated as clumps.  We also discuss the parameters for our clump finding algorithm and detail the clump detection process.  In Section~\ref{sec:sedfitting}, we summarize the derived properties of the host galaxies in which we find clumps, we derive stellar properties of clumps utilizing multi-band photometry for Spectral Energy Distribution (SED) fitting, and provide a comparison of statistical properties.  In Section~\ref{sec:intrinsicprop}, we discuss the physical properties of our sample including the number of clumps per host galaxy  and the size of clumps.  In Section~\ref{sec:discussion}, we investigate the relationship between sub-galactic clumps and their host galaxies by comparing the rest-frame 1500{\AA} UV flux and derived stellar properties of clumps to the overall properties of the host galaxies.  We also discuss any gradients that arise with respect to the galactocentric radius.  In Section~\ref{sec:conclusion}, we present a summary of our main findings.  In the Appendix (\ref{sec:appendix}), we quantify the effects of deriving clump properties with and without the inclusion of near-infrared (NIR) data as clumps are often not visible in the NIR.

Throughout this paper we assume cosmological parameters of $\Omega_{M} = 0.3$,  $\Omega_{\Lambda}  = 0.7$, $H_{0} = 70 $ $km s^{-1}$ $Mpc^{-1}$, and the AB magnitude system (Oke 1974).

\section{Data and Observations} \label{sec:data}

Ultraviolet imaging of the Hubble Ultra Deep Field (hereafter UVUDF) was an HST Cycle 19 program (HST PID 12534; PI: Teplitz) comprised of 90 orbits in total with the WFC3 UVIS detector in F225W, F275W, and F336W ($U$) filters over 3 epochs.  30 orbits per filter were obtained with a common pointing center, RA: 03\textsuperscript{\textit{h}}32\textsuperscript{\textit{m}}38.5471\textsuperscript{\textit{s}} DEC: -27\textsuperscript{$\circ$}46$'$59.$''$00 (J2000) and a pixel scale of 0.03$''$/pixel.  Decreasing charge transfer efficiency (CTE) caused by damage to the CCD lattice has resulted in the loss of data quality and is a problem for the imaging of faint sources.  We address this issue by using post-flashed data to mitigate the effects of CTE degradation \citep{Mackenty2012}.  Post-flash protects against the loss of the faintest objects by filling in``traps'' on the CCD before readout.  We use the post-flashed unbinned epoch-3 images of the three UV filters from the UVUDF program made 2012 August 3 - 2012 September 7 that consists of 16 orbits in F225W, 16 orbits in F275W, and 14 orbits in F336W.  These CTE corrected mosaics, which remove the affects of CTE on the observed morphology of the galaxies, were combined following the approaches described in \citet{Koekemoer2002, Koekemoer2011}.  A detailed description of data reduction and photometry can be found in \citet{Teplitz2013} and \citet{Rafelski2015}.

Additionally, observations from the Wide Field Camera (WFC) of the Advanced Camera for Surveys (ACS) provide the optical images used in our data set.  F435W($B$), F606W($V$), F775W($i$), and F850LP($z$) make up the optical filters used with pixel scale 0.03$''$/pixel.  These observations are from the ACS optical HUDF Cycle 12 program \citep{Beckwith2006}.  Details of the image processing and analysis are presented in \citet{Beckwith2006}.  Observations using WFC3 F105W($Y$), F125W($J$), F140W($JH$), and F160W($H$) filters comprise the set of infrared images used.  These data are from the HUDF09 (HST PID 11563; PI: Illingworth) and HUDF12 (HST PID 2498; PI: Ellis) programs.  Details of the image processing and analysis are presented in \citet{Koekemoer2013} and \citet{Ellis2013}; see also \citet{Illingworth2013}.  Table 1 from \citet{Rafelski2015} provides information for each bandpass including effective wavelength, zero point, exposure time, and depth. 

The UVUDF catalog from \citet{Rafelski2015} is used to select the target galaxy sample.  We apply two signal-to-noise (SN) cuts of 3$\sigma$ and 5$\sigma$ in F160W to remove spurious detections, which results in 1,404 and 1,200 galaxies respectively, with photometric redshifts of 0.5 $\leq z \leq$ 1.5.  Hereafter, we will be using the 3$\sigma$ cut catalog.  Galaxy selection and processing is discussed in further detail in Section~\ref{sec:detection}.

\section{Clumps} \label{sec:clumps}

\subsection{Clump Definition} \label{sec:definition}

\begin{figure}
\centering
\includegraphics[width=3.75cm]{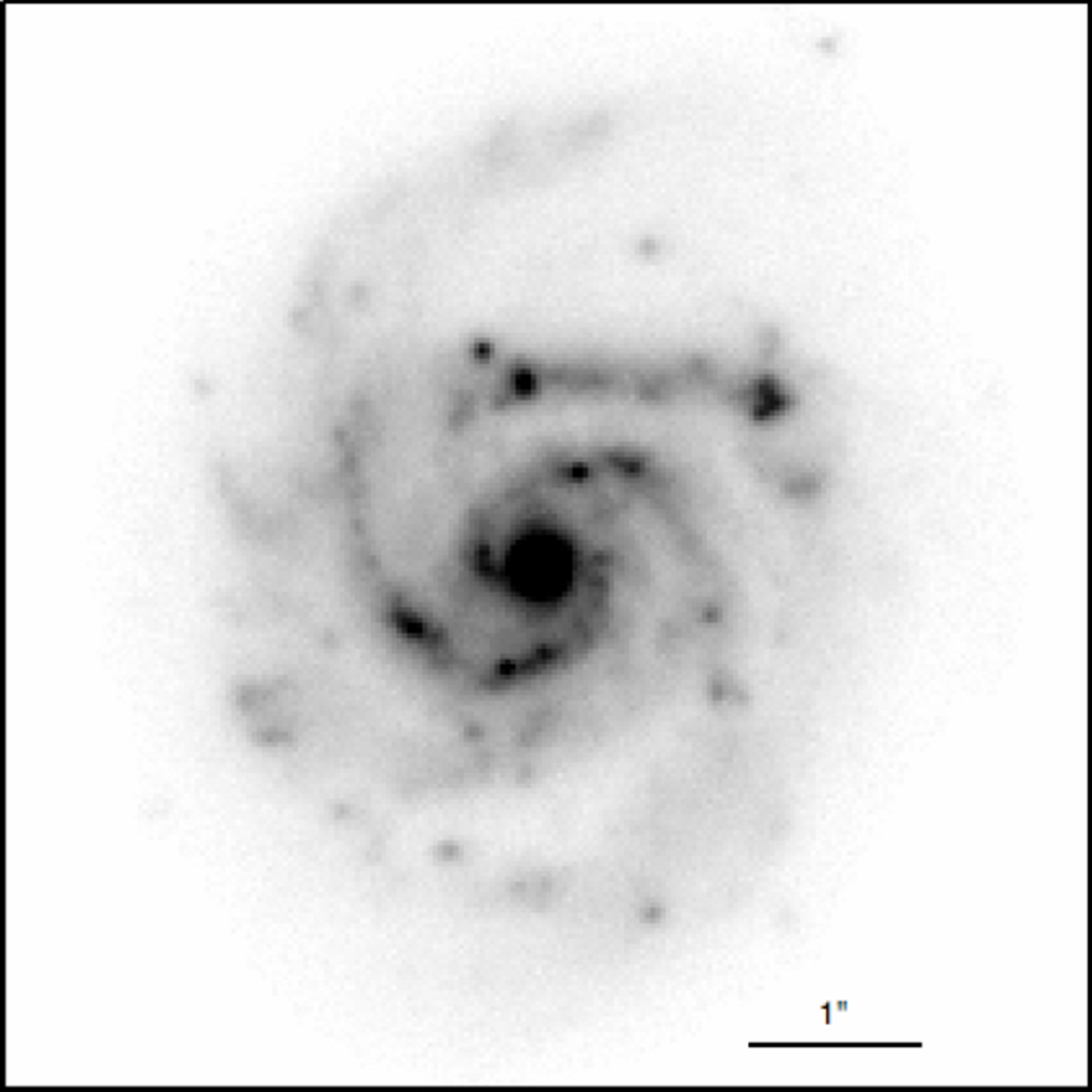}\includegraphics[width=3.75cm]{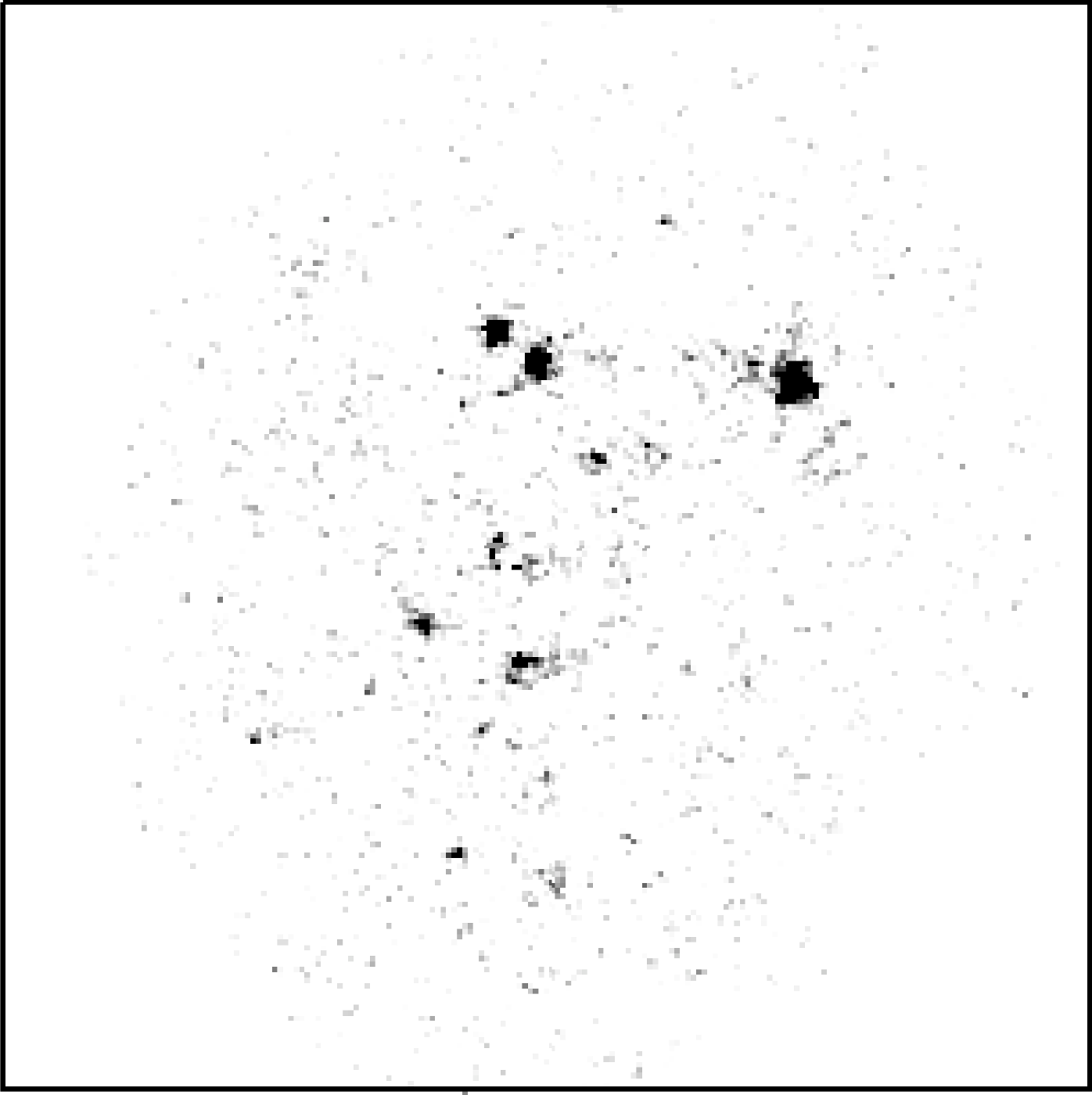}\\   
\includegraphics[width=3.75cm]{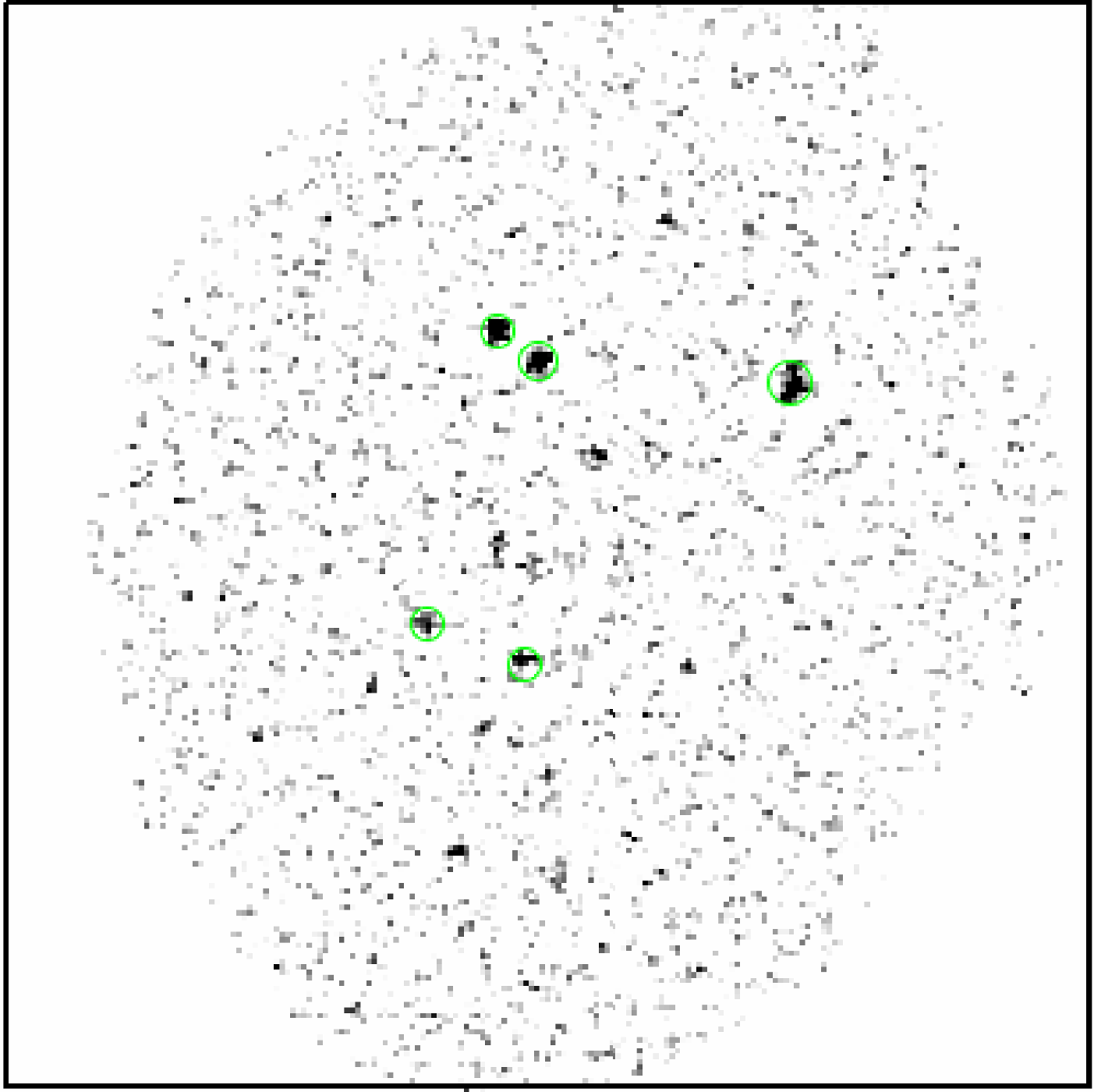}\includegraphics[width=3.75cm]{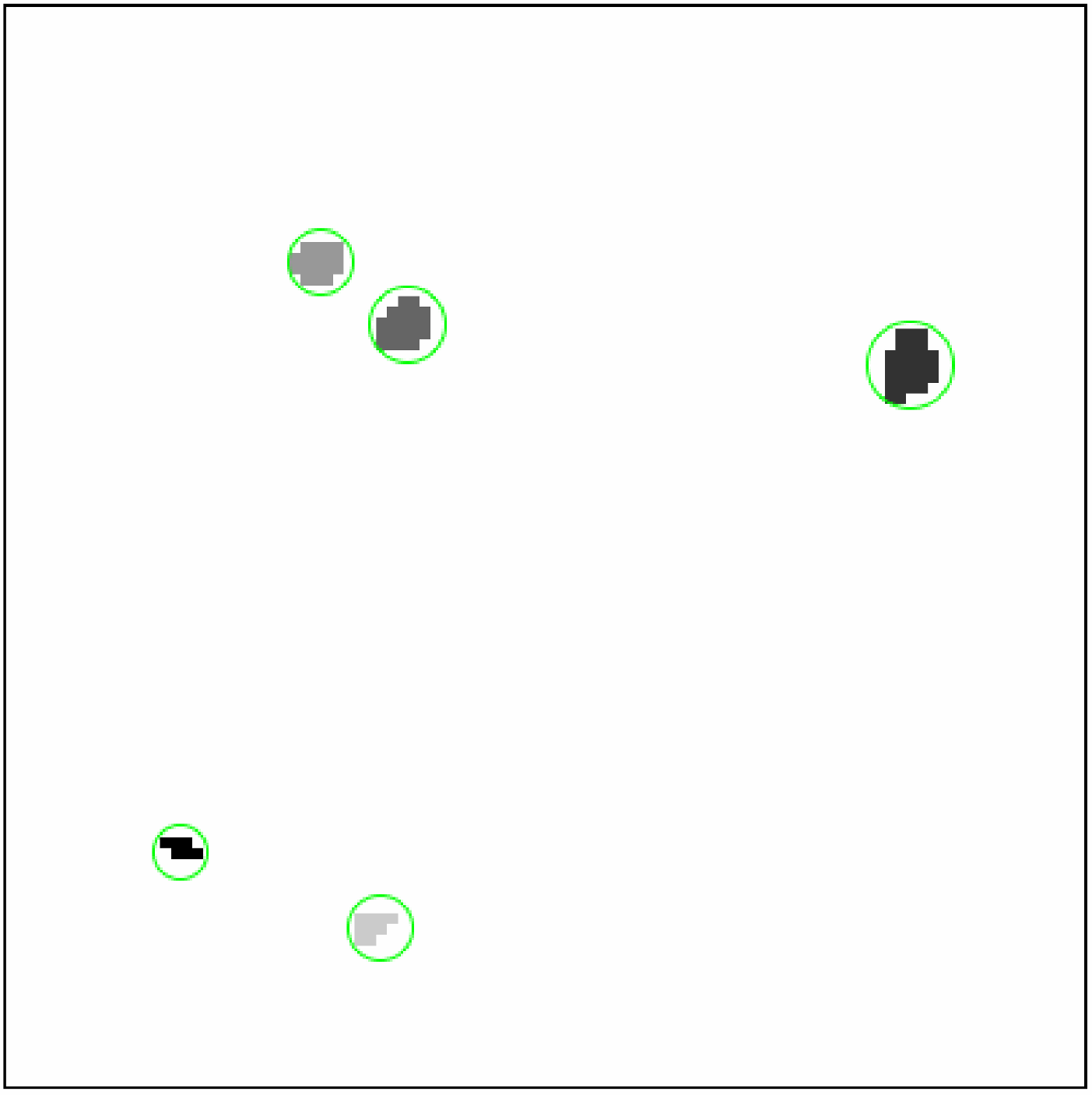}
\caption{Top left: F775W cutout image of a galaxy at $z=0.62$.  Horizontal line represents 1$''$ or 6.8 kpc. Top right: F225W image showing clumps (rest-frame UV).  Bottom left: contrast image created by subtracting the smoothed F225W image from the global background-subtracted F225W image.  Bottom right: zoomed-in image of the segmentation map created for the clumps detected.  Green circles show the detected clumps.
\label{clumpfinding}}
\end{figure}

One of the main difficulties in the study of clumps is how to define a clump.  Although we know more about clumps today than we did in the past, a globally accepted clump definition has yet to come into being.  Clumps may be defined loosely as clusters of hot stars \citep{Donahue2015}, or in a stricter fashion, as blobs whose UV luminosity is brighter than 8\% of the total UV luminosity of the galaxy in the NUV \citep{Guo2015}.  There are variations to this definition, such as that by \citet{Boada2015}, who define clumps as having UV luminosities $ > 1\%$ of the total galaxy UV light.  This alternative definition was used to include UV-faint clumps, since they are redder and would therefore provide greater insight into the Internal Color Dispersion (ICD) of their sample.  There are even more variations to the definition of a clump, such as \textit{hinge clumps} which are defined as luminous knots of star formation near the base of tidal features in interacting galaxies \citep{Smith2014}.

Although there are many clump definitions as detailed above, clumps are generally defined as small sub-galactic regions of intense star formation (often indicated by their brightness in the rest-frame UV, mainly in high redshift studies).  However, in this study we impose 4 additional constraints for such a region to be defined as a clump:

(i)  We automatically detect clumps using the rest-frame 1500{\AA} FUV light in the F225W, F275W, and F336W passbands.  A deficit of data at these wavelengths prevented previous use of the FUV for clump detection.  However, we attest that 1500{\AA} is a strong tracer of star formation and is ideal for clump detection especially at low and intermediate redshifts.

(ii)  We require that all visually identified clumps have 3$\sigma$ detection limit above the flux of the host galaxy in the detection band (rest-frame 1500{\AA} observed passband).

(iii)  Ensuring that galactic bulges are not mistakenly identified as clumps is also very important.  We therefore impose a 5 pixel minimum distance (a distance of 0.92-1.27 kpc for the redshift range selected) from the center of the galaxy to ensure that the clumps are not bulges, where the center of the galaxy is defined as the barycenter determined by SExtractor using the first order moments of the galaxy \citep{Bertin1996}.

(iv)  We also impose a minimum size limit based on images of local flocculent/clumpy galaxies, NGC 3521 and NGC 7331.  We use the physical sizes of local UV clumps (minimum of 0.46-0.64 kpc for the redshift range assuming a spherical geometry) to set the minimum size constraint of clumps at 0.5 $\leq z \leq$ 1.5.

We restrict our analysis to star-forming regions that meet the above criteria.  Clump detection and details on how the clump criteria were implemented are presented below where we discuss the clump detection algorithm.

\subsection{Clump Detection} \label{sec:detection}

We create a semi-automated clump finding algorithm to detect clumps and measure clump photometry, as illustrated in Figure~\ref{clumpfinding}.  We first create 9$''\times$9$''$ postage stamps of all $\sim$1,400 galaxies in our sample based on the coordinates generated from the primary SExtractor source detection run for the main UVUDF catalog performed in the \textit{B}-band \citep{Rafelski2015}.  We subtract the global background of the cutout and then use the $i$-band segmentation maps to isolate each galaxy to limit clump candidate detection to within the galaxy.  The postage stamps are then smoothed by applying a boxcar filter with a size of 10 pixels.  A contrast image is created by subtracting the smoothed image from the galaxy image.  From this we use simple image statistics to calculate the standard deviation of the contrast image in order to detect areas with possible clumps with SExtractor.  Based on the minimum size limit we set the limiting minimum area parameter in SExtractor to 5 pixels (from definition iv above, 5 pixels is approximately the minimum size at $0.5<z<1.5$).  We then run SExtractor with a 3{$\sigma$} detection limit per pixel (full description of SExtractor parameters used are listed in Table~\ref{sextractor}) on the contrast image to locate clump candidates.

\begin{deluxetable}{cc}
\tablecaption{SExtractor Input Parameters \label{sextractor}}
\tablecolumns{2}
\tablenum{1}
\tablewidth{0pt}
\tablehead{\colhead{Parameter} & \colhead{Value}}
\startdata
DETECT\_MINAREA & 5 pixels \\
THRESH\_TYPE & ABSOLUTE \\
DETECT\_THRESH & $3.0\sigma$ \\
ANALYSIS\_THRESH & $3.0\sigma$ \\
FILTER\_NAME & gauss\_3.0\_5X5.conv \\
DEBLEND\_NTHRESH & 32 \\
DEBLEND\_MINCONT & 0.0001 \\
CLEAN & Y \\
CLEAN\_PARAM & 5.0 \\
BACK\_SIZE & 32 \\
BACK\_FILTERSIZE & 3 \\
\enddata
\tablecomments{A full description of each SExtractor parameter can be found in \citet{Bertin1996}.}
\end{deluxetable}

We detect $\sim500$ clump candidates in 209 galaxies in one of three HST UV bands based on the photometric redshift of the galaxies at rest-frame 1500{\AA}, shown in Table~\ref{detectionband}.  This new UV HST data enables us to observe rest-frame 1500{\AA} for this redshift range for the first time.

\begin{deluxetable}{ccc}
\tablecaption{Detection Band \label{detectionband}}
\tablecolumns{3}
\tablenum{2}
\tablehead{\colhead{Redshift} & \colhead{Filter} & \colhead{Number of Galaxies}}
\startdata
$0.5 \leq z < 0.75$ & F225W & 57 \\
$0.75 \leq z < 1.0$ & F275W & 37 \\
$1.0 \leq z \leq 1.5$ & F336W & 115 \\
\enddata
\end{deluxetable}

Once the clump candidates are located in the contrast image, we proceed by measuring the flux in the detection band and in the remaining 6 filters (all of which are at the original HST resolution, FWHM $\sim$0.10$''$) utilizing SExtractor, subtracting the global background in each respective observed passband.  We also subtract the local background of the clump candidates in each filter which accounts for the galaxy background flux that contributes to the clumps.  This is accomplished by masking the clumps and then determining the median contribution from the rest of the galaxy.  To ensure that bulges are not included in the clump detection process, we require the condition that all clumps are at least 5 pixels away from the center of the galaxy.  The center of the galaxy was obtained from the catalog presented in \cite{Rafelski2015}.  The resulting clump catalog consists of 403 clumps detected in 209 host galaxies in the UVUDF.

Our clump finding algorithm is reminiscent of the automated star-forming region finder detailed in \citet{Guo2015} which also uses a contrast image for clump detection.  Their paper provides further details on the selection benefits of a 10 pixel boxcar filter for clump detection and the resultant limitations on the clump size this creates.  However, our procedure differs by one vital step, the detection bands in which the rest-frame UV light is measured.  Whereas the Guo et al. study was limited to detecting clumps in the observed optical bands, and therefore probing rest-frame 2200{\AA} and 2500{\AA} which include light from older stars, we are able to measure rest-frame 1500{\AA}, a strong tracer of the star formation from the younger stellar population \citep{Calzetti2013}, in their observed UV bands.  By looking further in the UV, we ensure that we are measuring the light from the youngest population and also minimize the contamination from older stars.

\section{SED Fitting} \label{sec:sedfitting}

\subsection{Host Galaxy Sample} \label{sec:hgsample}

\begin{figure}[ht!]
\epsscale{1.2}
\plotone{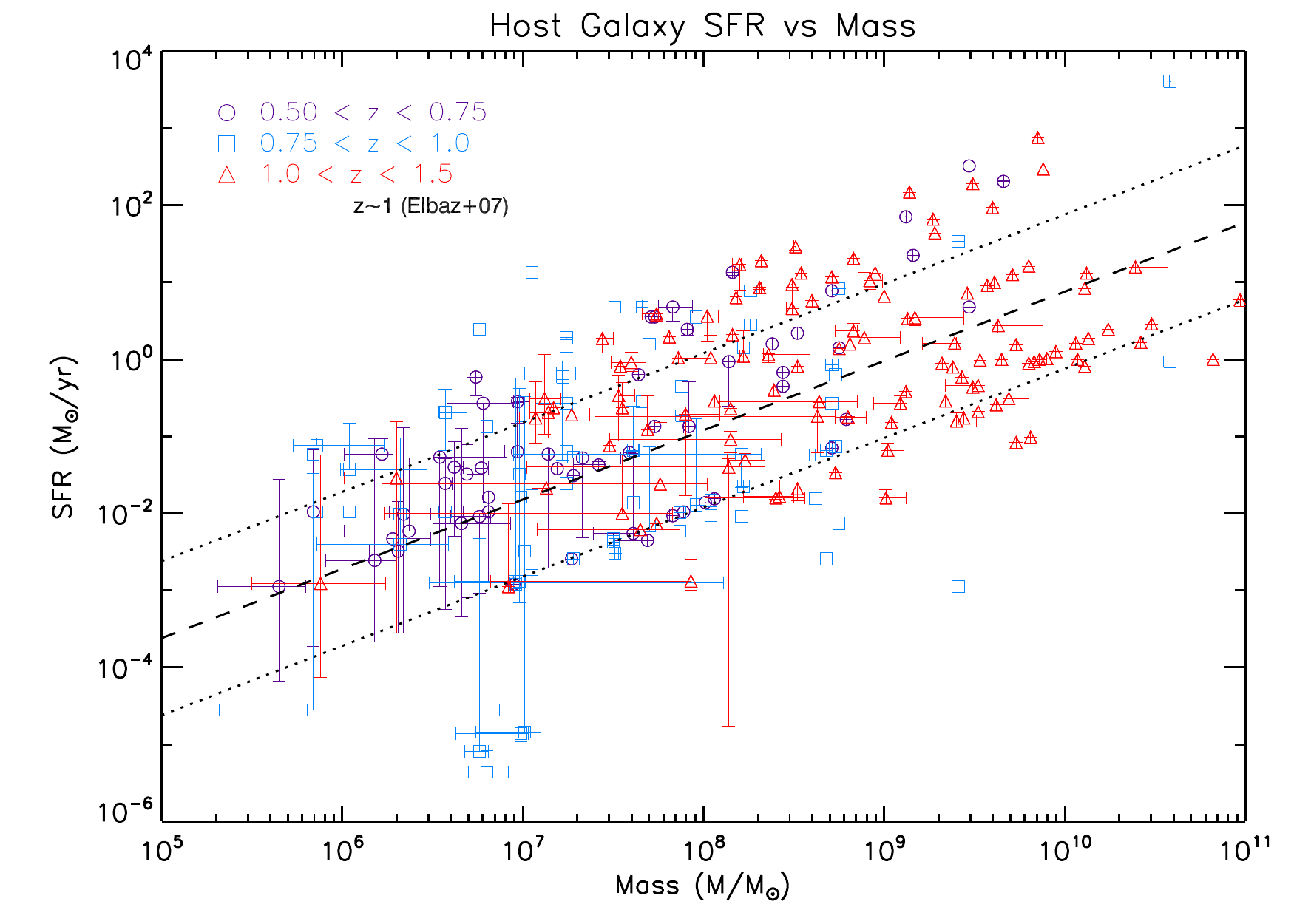}
\caption{SFR vs mass for host galaxies.  Log-Log plot with data points color-coded depending on photometric redshift, and spectroscopic redshift when available, as follows:  $0.5<z<0.75$ in purple circles, $0.75<z<1.0$ in blue squares, and $1.0<z<1.5$ in red triangles.  For comparison, the star-forming main sequence (SFMS) as determined by \citet{Elbaz2007} for a sample at $z\sim1$ is shown as a dashed line.  The dotted lines represent a 1.0 dex scatter from the Elbaz SFMS.  Error bars are from the confidence levels calibrated by FAST using Monte-Carlo simulations (see Section~\ref{sec:hgsample} for details).  For most cases error bars are too low and are not visible in the plot.
\label{hostgalsfrmass}}
\end{figure}

After applying the clump finding algorithm to the complete sample of galaxies, we create a subsample of 209 host galaxies that contain at least one clump which is not in the central region of the \textit{B}-band detection aperture.  We fit SEDs to the host galaxies using FAST on the multi-wavelength photometry from \citet{Rafelski2015} from the NUV through the NIR.  FAST \citep{Kriek2009} enables us to fit stellar population synthesis models from \citet{BC2003} assuming a \citet{Chabrier2003} IMF, Calzetti dust law \citep{Calzetti1994, Calzetti2000}, and exponentially declining SFR (EXP SFR) while constraining the redshift to the photometric redshifts of \citet{Rafelski2015} and spectroscopic redshifts when available.  From this we obtain galaxy properties including age, mass, and SFR.  The 1$\sigma$ (68\%) confidence levels are calibrated using 500 Monte Carlo simulations per galaxy and are described in the appendix of \cite{Kriek2009}.  From FAST we found that host galaxies have a median metallicity equal to $Z_{\odot}$, a median age of $\sim$25 million years, and a median SFR of $0.29_{-0.05}^{+0.15}$ M$_{\odot}$/yr (hereafter all given uncertainties are the 68\% confidence intervals).  They cover a broad range of stellar masses with a median mass of $1.66_{-0.25}^{+0.74}$ $\times$10\textsuperscript{8} M$_{\odot}$, agreeing well with Figure 27 from \citet{Skelton2014} that shows the evolution of mass as a function of redshift over H$_{F160W}$ AB mag bins.

It is well known that the SFR density evolves with redshift peaking at $z\sim2$ followed by a sharp decline \citep[e.g.][]{Bouwens2015, Madau2014, Cucciati2012}.  The rapid decrease in the SFR density in the redshift range 0.5-1.5 makes it an important redshift interval to study the assembly of galaxies.  The formation of a main sequence in the SFR versus stellar mass plane, which strongly evolves with redshift, is observed for galaxies at $0.2 < z < 2$ \citep{Speagle2014, Elbaz2007, Whitaker2012, Wisnioski2015, Fumagalli2014, Peng2010}.  The scatter in this relation could provide interesting constraints on the star formation history \citep{Kurczynski2016, Salmon2015, Shivaei2015}.  We show the SFR-mass relation for our host galaxy sample in Figure~\ref{hostgalsfrmass}, which shows that the galaxies follow the same increasing mass and SFR trend as that depicted by the star-forming main sequence (SFMS) determined by \citet{Elbaz2007} for galaxies at $z\sim1$.  More than half of the host galaxies lie above the SFMS shown and there are several galaxies which are beyond the 1.0 dex scatter (depicted as dotted lines in Fig.~\ref{hostgalsfrmass}) from the Elbaz SFMS.  We find that we have many star-forming galaxies which are typical main sequence galaxies but also about 56\% with SFR greater than depicted by the SFMS and 27\% greater than 1.0 dex above the SFMS.

\subsection{Clump Fitting with FAST} \label{sec:clumpfast}

\begin{deluxetable}{cc}
\tablecaption{FAST Input Parameters \label{fastinput}}
\tablecolumns{2}
\tablenum{3}
\tablehead{\colhead{Parameter} & \colhead{Value}}
\startdata
AB$_{-}$ZEROPOINT & 23.93 {$\mu$}Jy  \\
FILTERS$_{-}$RES & FILTER.RES.v7.R300 \\
N$_{-}$SIM & 0 \\
C$_{-}$INTERVAL & 68$\%$ \\
LIBRARY & bc03  \\
RESOLUTION & pr,lr\\
IMF & ch \\
SFH & exp,del \\
DUST$_{-}$LAW & calzetti \\
LOG$_{-}$TAU$_{-}$MIN & 6.5 log[yr] \\
LOG$_{-}$TAU$_{-}$MAX & 10.9 log[yr] \\
LOG$_{-}$TAU$_{-}$STEP & 0.2 log[yr] \\
LOG$_{-}$AGE$_{-}$MIN & 6.0 log[yr] \\
LOG$_{-}$AGE$_{-}$MAX & 10.1 log[yr] \\
LOG$_{-}$AGE$_{-}$STEP & 0.2 log[yr] \\
A$_{-}$V$_{-}$MIN & 0 mag \\
A$_{-}$V$_{-}$MAX & 3.0 mag \\
A$_{-}$V$_{-}$STEP & 0.1 mag \\
METAL & [0.004, 0.008, 0.020], [0.020] \\
H$_{0}$ & 70.0 \\
OMEGA$_{-}$M & 0.3 \\
OMEGA$_{-}$L & 0.7 \\
\enddata
\end{deluxetable}

Clump mass, metallicity, age, and SFR are determined using the same methodology as for the host galaxies for the 403 clumps in our sample.  Clumps at intermediate redshifts are often times not seen or resolved in the IR.  We consider the quality of the fits and the effects of not including the IR in the SED fitting in the appendix.  Based on these findings we do not include the IR data because of the resolution in the NIR.  Therefore, we limit our SEDs to the observed UV and optical photometry, and thus do not sample the older and/or low mass stellar populations.  We compare fits with an exponentially declining SFH (EXP SFH) and a delayed exponentially declining SFH (DEL SFH), both with minimum e-folding time of log$_{10}$($\tau$/yr) = 6.5.  It is important to fit the model with the most representative SFH to ensure that the Balmer break is properly sampled and thus provide accurate ages, and we discuss this further in Section~\ref{sec:cage}.

\citet{Torrey2014} find that a minimum of 5 bands covering a large wavelength range is required to obtain relatively good mass estimates as we do here with the filters in our study.  However, we are looking at star-forming regions and thus the clumps may only be detected in the UV or the blue optical.  This could further limit our sample; however, these cases are also very interesting because they are truly probing the birth of the clump, the youngest/hottest stars.  \citet{Torrey2014} also found that derived stellar mass errors can improve by constraining the metallicity and age range.  We use FAST with two sets of input parameters: (1) allowing the metallicity to float as an output parameter with Z = 0.20Z$_{\odot}$, 0.40Z$_{\odot}$, and Z$_{\odot}$, and (2) fixing the input to solar metallicity (Z = 0.020 = Z$_{\odot}$).  A full listing of the input parameters used for SED fitting are provided in Table~\ref{fastinput} and a listing of derived clump properties are provided in Table~\ref{clumprop}.

\begin{deluxetable*}{ccccc}
\tablecaption{Derived Clump Properties \label{clumprop}}
\tablecolumns{5}
\tablenum{4}
\tablehead{\colhead{Property} & \colhead{EXP$_{Z=Z_{\odot}}$} & \colhead{EXP$_{Z_{Float}}$} & \colhead{DEL$_{Z=Z_{\odot}}$} & \colhead{DEL$_{Z_{Float}}$} }
\startdata
\textbf{Mass (10\textsuperscript{7} M/M$_{\odot}$)} & & & &\\
Median & 0.65$^{+0.062}_{-0.083}$ & 0.63$^{+0.077}_{-0.094}$ & 0.71$^{+0.051}_{-0.077}$ & 0.68$^{+0.10}_{-0.059}$ \\
Average & 25.9$^{+13.6}_{-20.9}$ & 26.4$^{+13.6}_{-21.0}$ & 26.1$^{+13.6}_{-21.2}$ & 26.6$^{+13.7}_{-21.1}$ \\
\tableline
\textbf{Age (Myr)} & & & &\\
Median & 15.8$^{+24}_{-0}$ & 6.3$^{+0}_{-0}$ & 39.8$^{+23}_{-15}$ & 10.0$^{+0}_{-4}$ \\
Average & 344$^{+35}_{-36}$ & 356$^{+42}_{-42}$ & 407$^{+40}_{-45}$ & 390$^{+41}_{-40}$ \\
\tableline
\textbf{Star Formation Rate (M$_{\odot}$/yr)} & & & &\\
Median & 0.014$^{+0.006}_{-0.004}$ & 0.041$^{+0.019}_{-0.016}$ & 0.017$^{+0.004}_{-0.004}$ & 0.051$^{+0.040}_{-0.018}$ \\
Average & 4.43$^{+0.66}_{-0.64}$ & 6.40$^{+0.85}_{-0.82}$ & 10.40$^{+1.57}_{-1.58}$ & 12.0$^{+1.49}_{-1.38}$ \\
\enddata
\tablecomments{Values above are shown for EXP (Exponentially Declining SFH) and DEL (Delayed Exponentially Declining SFH)\\
	Metallicity parameters for data presented are as follows: Z$_{\odot}$ (solar metallicity) and Z$_{Float}$=[0.20Z$_{\odot}$, 0.40Z$_{\odot}$, Z$_{\odot}$] (floating metallicity)\\
	The 68\% confidence limits for the averages and medians are denoted for each value in the table.}
\end{deluxetable*}

\subsection{Metallicity and A$_V$} \label{sec:metal}

Other clump studies often fix clump metallicity in their models such as \citet{Wuyts2013} who assume clumps have solar metallicities and \citet{Elmegreen2009a} who confine their studies to clumps of Z = 0.4Z$_{\odot}$.  These studies assume particular parameters beforehand in order to further constrain their models.  We investigate two cases: (1) where Z = 0.20Z$_{\odot}$, 0.40Z$_{\odot}$, and Z$_{\odot}$ and (2) when fixing the clumps with Z = Z$_{\odot}$, while constraining the redshift for both cases.  The average extinction for clumps with floating metallicity is $0.97_{-0.03}^{+0.04}$ mag and $0.98_{-0.03}^{+0.03}$ mag for EXP and DEL SFH respectively, where the median for EXP SFH is $0.90_{-0.10}^{+0.10}$ mag and $1.00_{-0.10}^{+0.00}$ mag for DEL SFH.  Similarly, the average extinction for these SFH models when constraining to solar metallicity are very similar with extinctions of $0.86_{-0.03}^{+0.03}$ mag and $0.87_{-0.03}^{+0.03}$ mag respectively, and a median of $0.70\pm0.10$ mag for both models.

We perform Kolmogorov-Smirnov (KS) tests on the physical properties for each model to quantify how the output parameters compare with one another.  The 2-sample KS test uses a cumulative distribution function to estimate the probability, P$_{KS}$, that both samples are drawn from the same parent distribution.  Low significance levels indicate that the two data sets are significantly different, while high values indicate they are probably consistent with a single distribution.  Tests show that the distribution of the extinction ($A_{V}$) between the two SFHs are statistically the same with P$_{KS} = 1.00$ for both floating and solar metallicity.  Through visual inspection we find that the model with constrained solar metallicity and an exponentially declining star formation history has the best fit SEDs with a median SFR of 0.014 M$_{\odot}/$yr.  In comparison, the median SFR of the host galaxies is 0.29 M$_{\odot}/$yr, less than the SFR of our own Milky Way galaxy (SFR = 0.68 - 1.45 M$_{\odot}/$yr; \citealt{Robitaille2010}).

\subsection{Mass} \label{sec:cmass}

\begin{figure}[ht!]
\epsscale{1.15}
\plotone{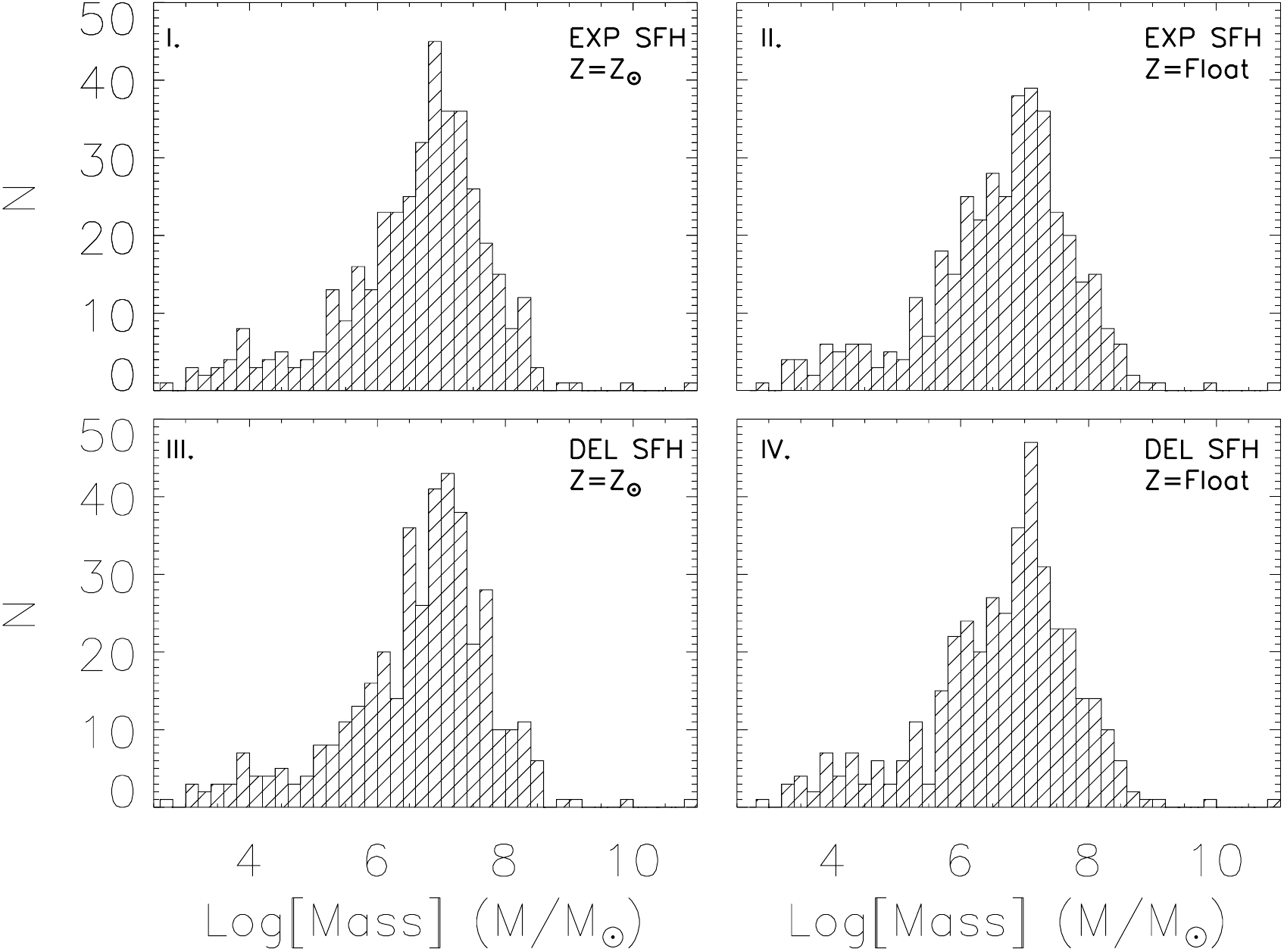}
\caption{Mass distribution of clumps for two SFHs (bin size = 0.20) and two metallicities.  Top left (I): exponentially declining SFH (EXP) and solar metallicity.  Top right (II): exponentially declining SFH (EXP) and floating metallicity,  Bottom left (III): delayed exponentially declining SFH (DEL) with solar metallicity.  Bottom right (IV): delayed exponentially declining SFH (DEL) with floating metallicity.
- Figures~\ref{massdist} and~\ref{agedist} are bin size = 0.20 and are organized in the same manner as described above for the two SFH and metallicity models.
 \label{massdist}}
\end{figure}

Mass distribution histograms (Fig.~\ref{massdist}) show that regardless of the assumed SFH, EXP or DEL, the distribution of mass for all the clumps agree with one another with P$_{KS} > 0.90$ for all models.  The histograms show a distribution peak at 1.0 $\times$ 10\textsuperscript{7} M$_{\odot}$ and clump mass range primarily from 10\textsuperscript{3} M$_{\odot}<$ $M_{clump}<10\textsuperscript{9}$ M$_{\odot}$.  Similarly, \citet{Elmegreen2005-apr} find clump masses ranging from 10\textsuperscript{6} M$_{\odot}$ to 10\textsuperscript{8} M$_{\odot}$ for galaxies with masses from 10\textsuperscript{9} M$_{\odot}$ to 10\textsuperscript{11} M$_{\odot}$.  Mass results from \citet{Elmegreen2013} for UDF, Kiso, and local galaxy clumps span a mass range of 10\textsuperscript{$\sim$3-9} M$_{\odot}$, where higher mass clumps were found in the UDF sample and lower mass clumps in both the Kiso and local galaxy sample.  The masses of clumps from our sample are in agreement with those determined from simulations \citep{Tamburello2015} and observational studies \citep{Livermore2015, Adamo2013} at similar redshifts and at $z>2$ \citep{Swinbank2009}.  Although clump masses for our sample cover a wide range, they are consistent with masses determined by other clump studies.

One important aspect to take into consideration are the masses determined for clumps in the high redshift bin.  For those with $z > 1.15$, the only filter sampling the rest-frame optical and 4000{\AA} break is F850LP, with no additional filter red-ward of this.  There are a total of 233 clumps in the high redshift bin and 108 of these are at $z > 1.15$.  The precise position of the break is known for clumps with spectroscopic redshifts ($\sim$ 30\%); however, this is not the case for the remaining clumps with only photometric redshifts.  For these clumps there may be very little constraint on the stellar masses determined; however, we retain them as part of the high redshift bin sample.  We distinguish between the two subsamples ($1.0<z<1.15$, $1.15<z<1.5$) in the high redshift bin with filled and open symbols.

\subsection{Age} \label{sec:cage}

\begin{figure}[ht!]
\epsscale{1.15}
\plotone{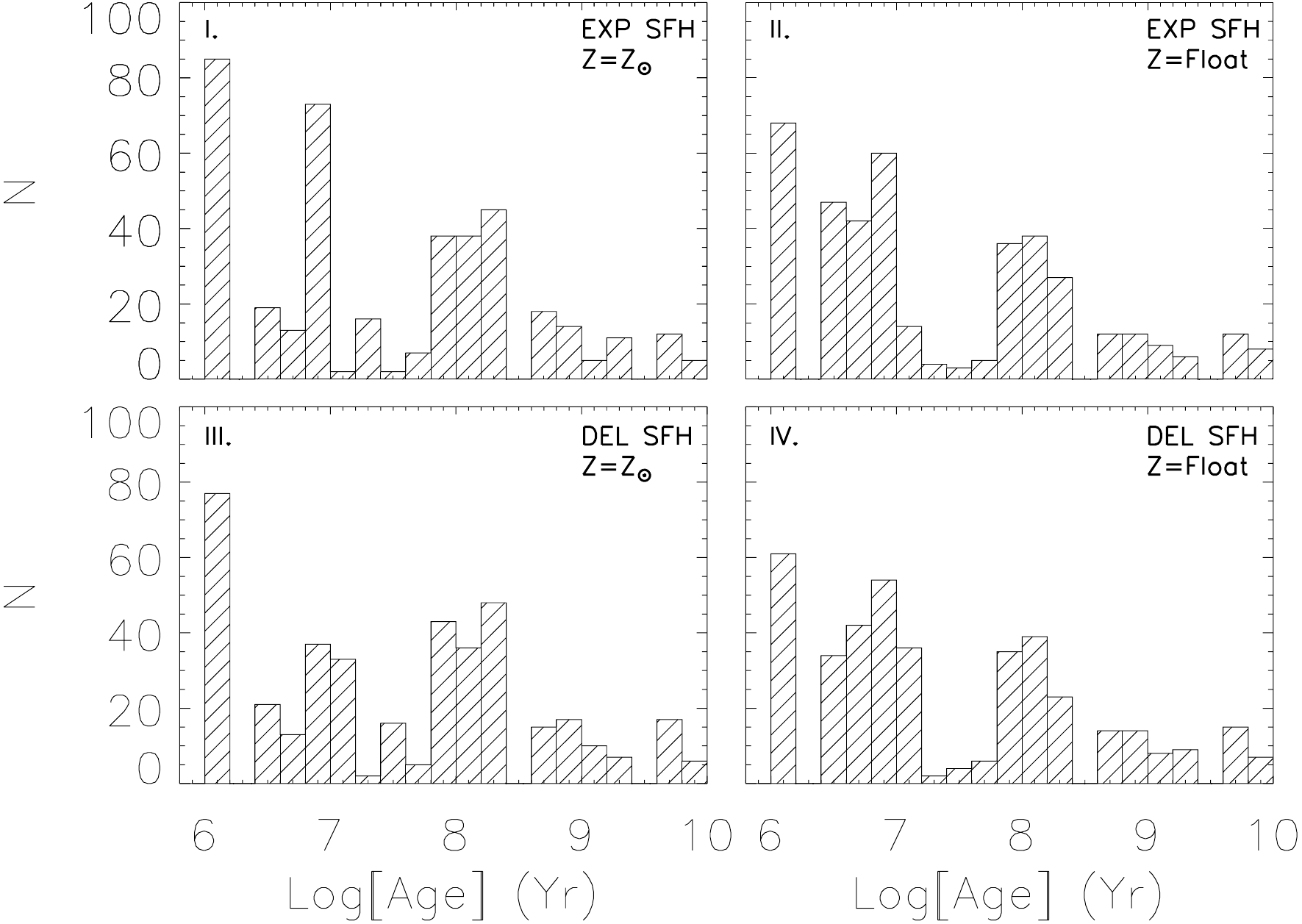}
\caption{Plot of the distribution of the age of clumps.  The ages for both SFH and metallicity models display a similar distribution with peaks at 10\textsuperscript{6}, 10\textsuperscript{7}, and 10\textsuperscript{8} yrs.  See Fig. \ref{massdist} for caption explanation.
\label{agedist}}
\end{figure}

As shown in Figure~\ref{agedist}, the distribution of clump ages for both EXP and DEL SFH agree rather well.  The distributions peak at 10\textsuperscript{6} yr with consistently smaller peaks at $\sim$10\textsuperscript{7} yr and $\sim$10\textsuperscript{8} yr.  Although the maximum deviation between the cumulative distribution, the KS statistic, is relatively low for solar metallicity (0.104) and floating metallicity (0.065), the corresponding significance levels indicate that they are less likely to be drawn from the same parent distribution.  The average age for the SFH models lies between 344 Myr - 407 Myr, where the EXP models tend toward the lower end of the range and DEL models toward the higher end.  The median age for each model is a few million years from about 6 to 40 Myr.   \citet{Elmegreen2013} found average ages for their clump sample: 12.6 Myr for the Kiso Clumps and 63.1 Myr for the UDF clumps which coincide with our median age values.  The average age of our clumps are higher by a factor of 10 as a result of a few clumps whose ages drive the average up.  However, clumps of this age range have been found at $z \sim 2$ \citep{Forster2011-sep2,Wuyts2012} which indicate that these older clumps are not completely uncommon.  The ``clumpy disk'' phase can last for several $10^{8}$ yr \citep{Jones2010} which may account for the older ages.

Some of our clumps may be older because, although they are relatively UV bright, there might be an underlying first generation stellar population present in the clump region.  The average age of the host galaxies is $1.04^{+0.09}_{-0.10}$ Gyr with a range of 10\textsuperscript{7} - 10\textsuperscript{9.8} yrs.  In Figure~\ref{agedist} we observe that most clumps are on the order of millions of years old; however, there are about 30 clumps which are much older (ages greater than 10\textsuperscript{9} yrs) that are about the same age as the host galaxies.  This implies that they may have formed after or at the same time as the host galaxy formed and could therefore be examples of the long lived clumps described by \cite{Ceverino2010} and \cite{Bournaud2007}.

\subsection{Star Formation Rate} \label{sec:csfr}

Clumps are known to form in galaxies with active star formation \citep[e.g.][]{Reddy2008, Daddi2007} and we find that the host galaxies have median SFR$_{UV+optical+IR}$ of {0.29} M$_{\odot}$/yr.  The peak of the SFR distribution for clumps ranges from 10\textsuperscript{-3} to 10\textsuperscript{-1} M$_{\odot}$/yr for all SED models with a median SFR of $0.014^{+0.006}_{-0.004}$ M$_{\odot}$/yr for EXP SFH and solar metallicity.  However, there are at least 35 clumps in each model displaying high SFRs (SFR $>$ 10 M$_{\odot}$/yr).  These highly star-forming clumps tend to be very young, corresponding to ages of about 1 million years, and with masses about one order of magnitude greater than the median mass of clumps, consistently for all SED models.  This corresponds to the fact that they are comprised of bright, young O \& B stars that are very massive themselves and could indicate an initial starburst accounting for the young ages and highly driven star formation.  \citet{Elmegreen2013} obtained SFRs for their clump sample and found a range of $10\textsuperscript{-4}-10\textsuperscript{2}$ M$_{\odot}$/yr which agrees quite well with the SFR range of the majority of the clumps.  Similarly, \citet{Livermore2012} found H$_{\alpha}$ SFRs with a range of $10\textsuperscript{-3}-10\textsuperscript{1}$ M$_{\odot}$/yr for clumps at $z=1-1.5$ which coincide with the SFRs of our high redshift bin clumps.
 
The clump sample has a few outliers with very low SFRs (SFR $<$ 10$^{-5}$ M$_{\odot}$/yr) which brings to question whether FAST chose the proper fit.  We visually inspect the clumps to determine the cause of the extremely low SFRs.  One of the factors that may be contributing to this is the extinction determined by FAST.  For example, the host galaxy for one of the outliers exhibits a lot of dust surrounding the clumps from an edge-on disk; however, FAST found no extinction correction.  FAST also found no extinction correction for a small face-on spiral with undefined arms.  Clearly these outliers are a product of the fits determined by FAST when a poor job of calculating the extinction occurs.  Although there are some clumps with low SFRs, we confirm that the host galaxy SFRs for these outliers range from 0.45-12.6 M$_{\odot}$/yr.  Therefore, regardless of this relationship we do find that our star formation rates agree well with the star-forming galaxies on the main sequence depicted as a black line in Figure~\ref{hostgalsfrmass}.

\section{Intrinsic Properties} \label{sec:intrinsicprop}

\subsection{Clump Number} \label{sec:cnumber}
 
\begin{figure}[ht!]
\epsscale{1.2}
\plotone{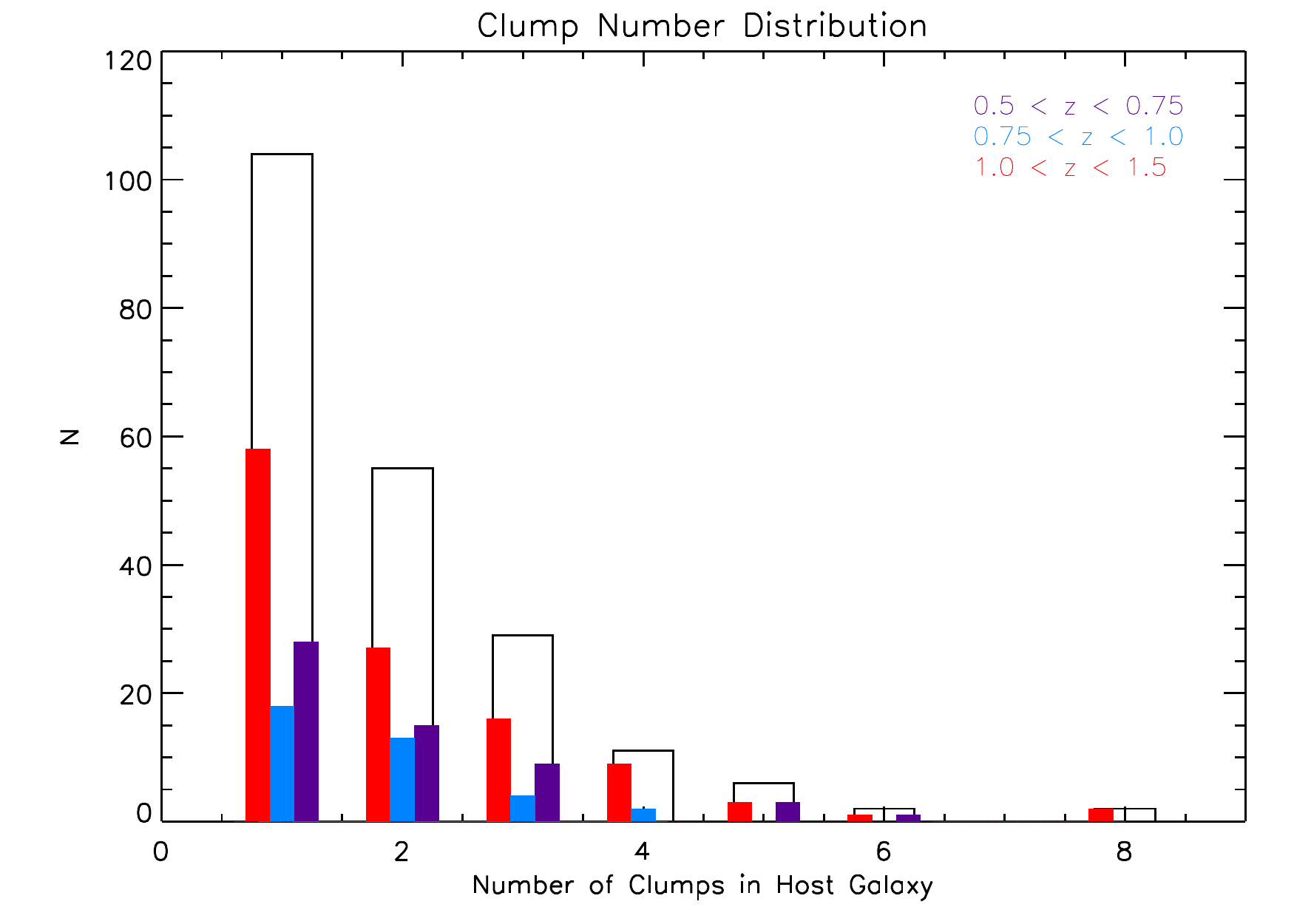}
\caption{Distribution of the number of clumps in host galaxies.  The black histogram represents all the clumps in the sample, the purple for clumps in the $0.5 < z < 0.75$ bin, the blue for clumps in the $0.75 < z < 1.0$ bin, and the red for clumps in the $1.0 < z < 1.5$ bin.
\label{clumpnum}}
\end{figure}

The number of clumps detected in each host galaxy is of great importance in understanding how the UV flux is distributed and can contribute toward an explanation for galaxy evolution.  Do small clumps form in the outskirts and over time migrate toward the center, in the process merging with other clumps to form fewer but brighter clumps?  Do we find that the galaxies with few clumps have them closer to the center?  Correlations between clump number and these other properties could be detectable with the use of the deepest UV imaging to date of the UDF, which pushes the limitations of previous clump detection even further.  We are likely to see faint galaxies that have never before been detected with the possibility that these have faint, previously unidentified clumps.  Figure~\ref{clumpnum} plots the distribution of the number of clumps found in galaxies.  The sample is dominated by single clump host galaxies and then declines as the number of clumps in the host galaxy increases.  We find that half of our host galaxies have 2 or more clumps, with the average number of clumps being 2.  This supports the findings of the simulations from \citet{Mandelker2014} that average $\sim2$ in-situ clumps per disk.  This indicates that the formation of a clump is not a singular event if clumps are caused by disk instabilities.  The instabilities which lead to their formation are drastic enough to cause multiple disturbances and thus create multiple clumps.

\subsection{Clump Size} \label{sec:csize}

\begin{figure}[ht!]
\epsscale{1.21}
\centering
\plotone{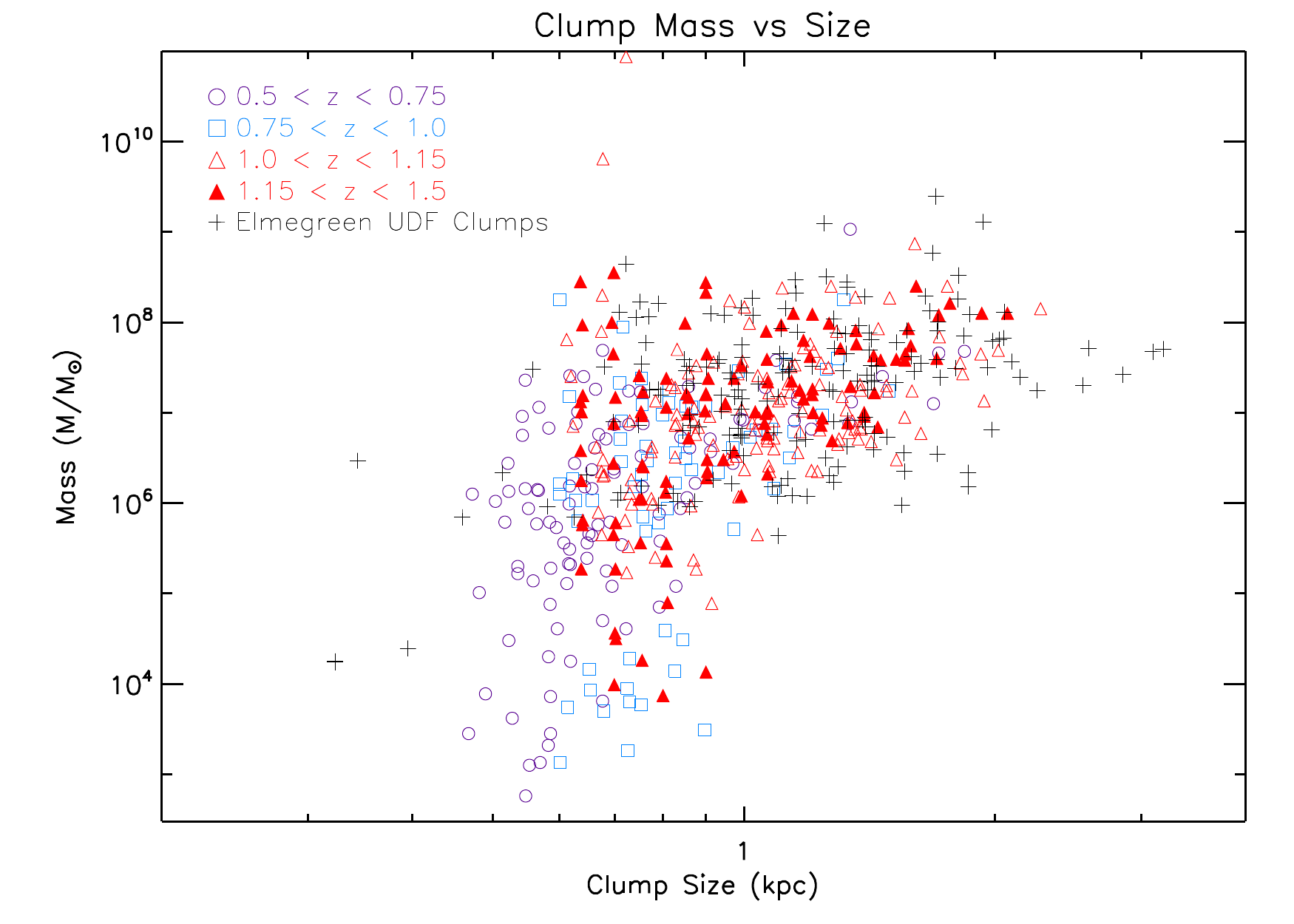}
\caption{Clump size vs mass plot for an exponentially declining SFH and solar metallicity SED fit. Purple circles are low redshift bin ($0.5<z<0.75$),  blue squares are intermediate redshifts ($0.75<z<1.0$), and red triangles are high redshift clumps (open - $1.0<z<1.15$, filled - $1.15<z<1.5$).  The black crosses represent the UDF clump sample from \citet{Elmegreen2013} at $z<3.6$.  Typical errors are given in Table~\ref{clumprop}.\\
- Redshift bins are color-coded in the same manner for Figures~\ref{massage} and~\ref{massratiodist}.
\label{sizemass}}
\end{figure}

Clump sizes were estimated for all 403 clumps by taking the pixel area of the clump in the detection band from the SExtractor output ISO\_AREA, with the smallest geometries meeting the resolution of HST.  With a pixel scale of 0.03$''$/pixel and PSF FWHM of 0.11$''$, 0.10$''$, and 0.09$''$ for F225W, F275W, and F336W respectively, we can measure clump diameters down to about 3.7 pixels in F225W and 3 pixels in F336W.  These sizes correspond to minimum diameters of 0.68 kpc in the lower redshift bin and 0.72 kpc in the higher redshift bins.  To automate the clump size determination we assume a spherical geometry for all clumps with diameter = 2 $\times$$\surd$(Area/$\pi$).  This is why in some cases clump size values may seem smaller than HST resolution.  We find clump diameters which are typically 3 to 5 pixels in the 3 redshift bins \citep{Elmegreen2009a}.  From these diameters and the photometric redshifts we determine the clump size in kpc.  Figure~\ref{sizemass} shows kpc-sized clumps which are comparable to clumps associated with all types of star-forming galaxies at $z\sim2$ \citep[e.g.][]{Genel2012-jan, Genzel2008, Genzel2011, Bournaud2008} and previous visual measurements performed on epoch 2 data of the UVUDF.  \citet{Elmegreen2013} found clump sizes $> 0.5$ kpc in UDF clumps and multiple giant clumps of $\sim1$ kpc for massive star-forming galaxies at $z\sim2$ \citep{Elmegreen2004, EE2005, Forster2006, Forster2011-sep2}.  \citet{Wisnioski2012} found a larger average clump size of $\sim$1.5 kpc for clumps at z$\sim$1.3.  Our data have an average clump size of 0.9 kpc but extend up to $\sim2$ kpc which are quite large relative to these studies and may be due to groups of nearby clumps which were not deblended.  The average clump size based on redshift bin are as follows: 0.75 kpc  ($0.5 < z < 0.75$), 0.84 kpc ($0.75 < z < 1.0$), and 1.06 kpc ($1.0 < z < 1.5$).  \citet{Elmegreen2009b} found that there was a general evolution toward smaller clumps and smoother disks at lower redshifts.  This correlation is seen here in the size averages per redshift bin where the difference between the lowest and highest redshift bins is 0.31 kpc.  However, this may in fact be an artificial trend created by the adopted minimum-area clump criterion since clumps in the higher redshift bin would be larger in order to meet the 5 pixel minimum.

Clump size increases in correlation to clump mass as shown in Figure~\ref{sizemass}.  This relationship was also observed in \citet{Elmegreen2013} for UDF clumps up to a redshift of $z < 3.6$.  We plot their data (black crosses) for comparison in Figure~\ref{sizemass}.  Additionally, we find an evolution of mass and size with respect to redshift, where lower redshift clumps comprise the smaller less massive clumps and higher redshift clumps dominate the larger and more massive end.  Many clumps at $0.5 < z < 1.0$ are at masses lower than those typically determined for the UDF clumps (less than about $10^{6}$ M$_{\odot}$) and of smaller size.  We attribute these lower masses and smaller sizes to our photometric data.  The UV coverage of the clumps provides us with the rest-frame FUV for our clump sample SED which enables us to determine these low masses for the lower redshift clumps.  However, it is also possible that this correlation may arise from Malmquist biases created by our limiting magnitude of the host galaxy sample.

\section{Discussion} \label{sec:discussion}

\subsection{Clumps vs. Galaxies} \label{sec:clumpvgal}

\begin{figure}[ht!]
\epsscale{1.2}
\plotone{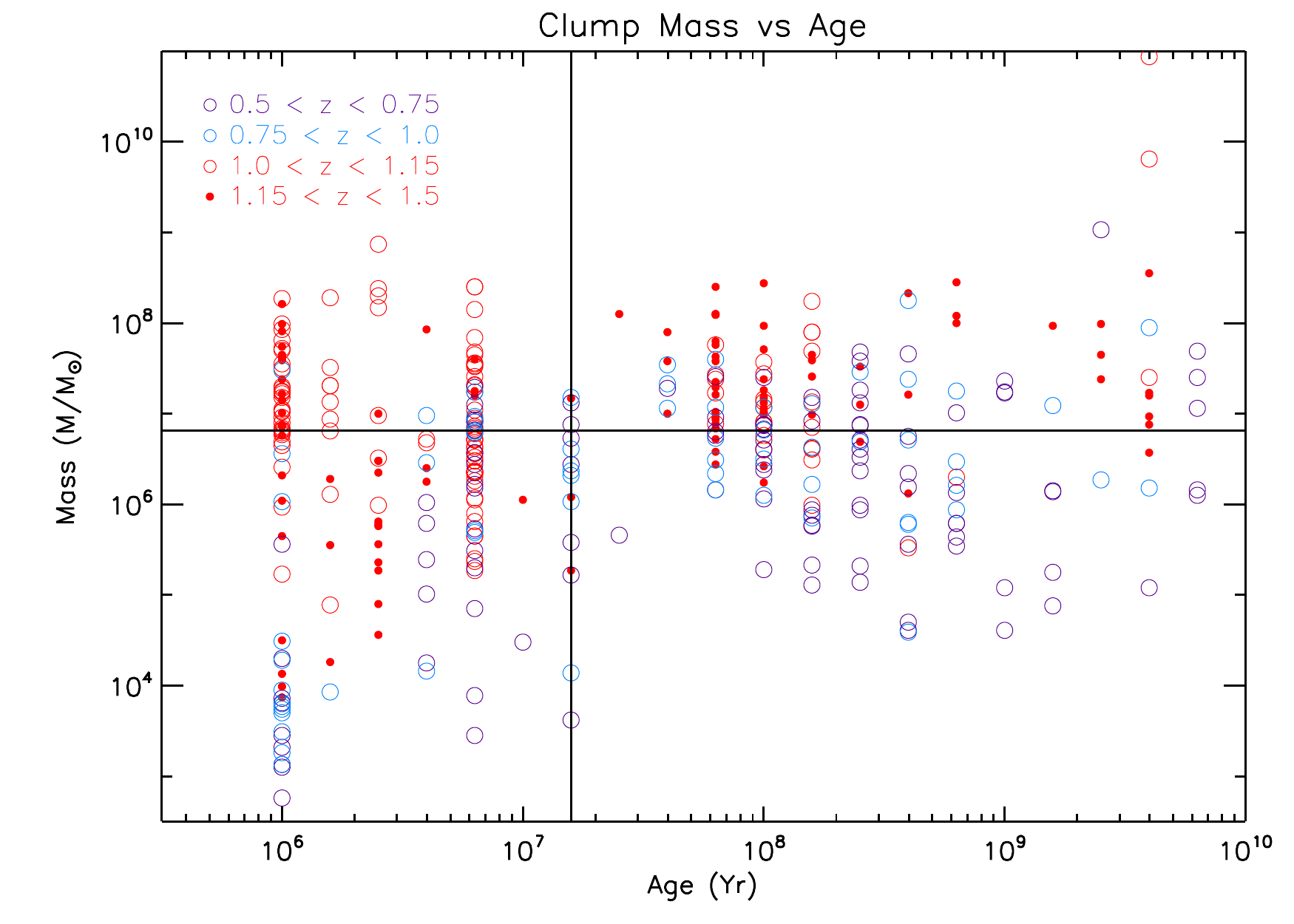}
\caption{Mass vs age plot of clumps.  Data points are color-coded based on redshift range as follows: open circles in purple for $0.5<z<0.75$, in blue for $0.75<z<1.0$, in red for $1.0<z<1.15$, and filled red circles for $1.15<z<1.5$.  Typical errors are listed in Table~\ref{clumprop}.  Horizontal and vertical solid black lines are the median mass (6.46$\times10\textsuperscript{6}$ M$_{\odot}$) and median age (15.8 Myr) of the clumps.
\label{massage}}
\end{figure}

Comparing the physical properties of clumps to the overall properties of their host galaxies is a crucial step in understanding the role that these sub-galactic regions truly play in the evolution of their hosts.  This can provide insights into clump migration, bulge formation, disk formation, and so forth.  For example, Figure~\ref{massage} shows that higher redshift galaxies are comprised of younger clumps ($< 10^{7.2}$ yr) that cover a broad range of masses but mostly dominate the higher mass end ($> 10^{6.8}$ M$_{\odot}$).  About 52\% of all clumps are young and of these 42\% are of greater mass than the median mass shown in Figure~\ref{massage}.  The concentration of high redshift clumps at young ages and large masses, which constitute about 18\% of all clumps in the sample, coupled with the highly star forming clumps mentioned in section~\ref{sec:csfr} could indicate newly formed clumps.  \citet{Tamburello2015} states that feedback suppresses clump formation and that massive clumps ($>10^{8}$ M$_{\odot}$) may only form via clump-clump mergers.  In simulations without feedback, they found that vigorous star formation gives rise to longer-lasting clumps that can reach very high masses before sinking in towards the bulge.  We find older, massive clumps which could be examples of these mergers.  Star formation rates, mass fractions, and flux ratios are indicators of these evolutionary stages.  Here we take a closer look at said properties for our clump sample and their host galaxies.

\subsubsection{Star Formation Rate} \label{sec:cgsfr}

Figure~\ref{hostgalsfrmass} shows the SFR as a function of mass and redshift, and that host galaxies have SFRs ranging on scales of 0.001 to 100s M$_{\odot}$/yr.  The SFRs of the host galaxies follow the increasing trend of the SFR and mass relationship depicted by the star-forming main sequence and are therefore representative of typical star-forming galaxies.  The clumps agree with the trend depicted by the SFMS as well; however, a large fraction of the clump sample lies well above ($\sim 1.0$ dex) the SFMS.  

\citet{Wuyts2013} found that fractional contribution of clumps to the integrated SFR of the star-forming galaxies increases to $\sim$ 20\% at $z\sim2$ \citep{Genel2012-jan, Genzel2011, Forster2011-sep2} but we do not see this trend toward a higher fractional contribution.  Studies by \citet{Guo2015} and \citet{Wisnioski2012} state that clumps individually contribute 4\%-10\% of the star formation.  According to the data, each clump generally contributes only a small fraction of the total star formation rate of the whole galaxy.  We find a broad range of fractional contributions, with a median contribution of about 5\% for each clump individually, which implies that the bulk of the the star formation is not in clumps.

\citet{Forster2011-sep2} suggested that the duration of the SFR activity is shorter for localized kpc-sized clumps compared to the bulk of the stellar population across the galaxy.  This would imply that most clumps would still be younger than the interclump regions, and our data supports this conclusion.  It would be reasonable to assume that the time for peak star formation within a clump would be much shorter considering that clumps are such a compact region in comparison to the total area of the galaxy and that the SFR would therefore be dominated by activity in other larger areas.

\subsubsection{Mass Fraction} \label{sec:massfrac}

Mass fraction is the fractional contribution of the clump mass to the total mass of the host galaxy, and is used as an indicator of clump interaction and migration toward the center of the galaxy for bulge formation.  If multiple clumps are present, it is possible that they may interact with each other, eventually losing angular momentum, and thus spiral into the center of the galaxy \citep[e.g.][]{Mandelker2014, Bournaud2007, Elmegreen2009b}.  \citet{Elmegreen2009b} estimated the ratio of clump mass to host galaxy mass, as we do here, and compared this with ratios in simulations that resulted in clumps migrating to the center of the galaxy.  They found that the total clump mass was $\sim$ 30\% of the disk mass where each clump was $\sim$ 5\% of the disk mass.  \citet{Ceverino2010} found lower estimates for the clump mass contribution where clumps comprised about $10-20\%$ of the total mass of the disk and \citet{Bournaud2014} found stellar mass fractions of 18\% for clumps ranging in age from 100-200 Myr.

\begin{figure}[ht!]
\epsscale{1.2}
\plotone{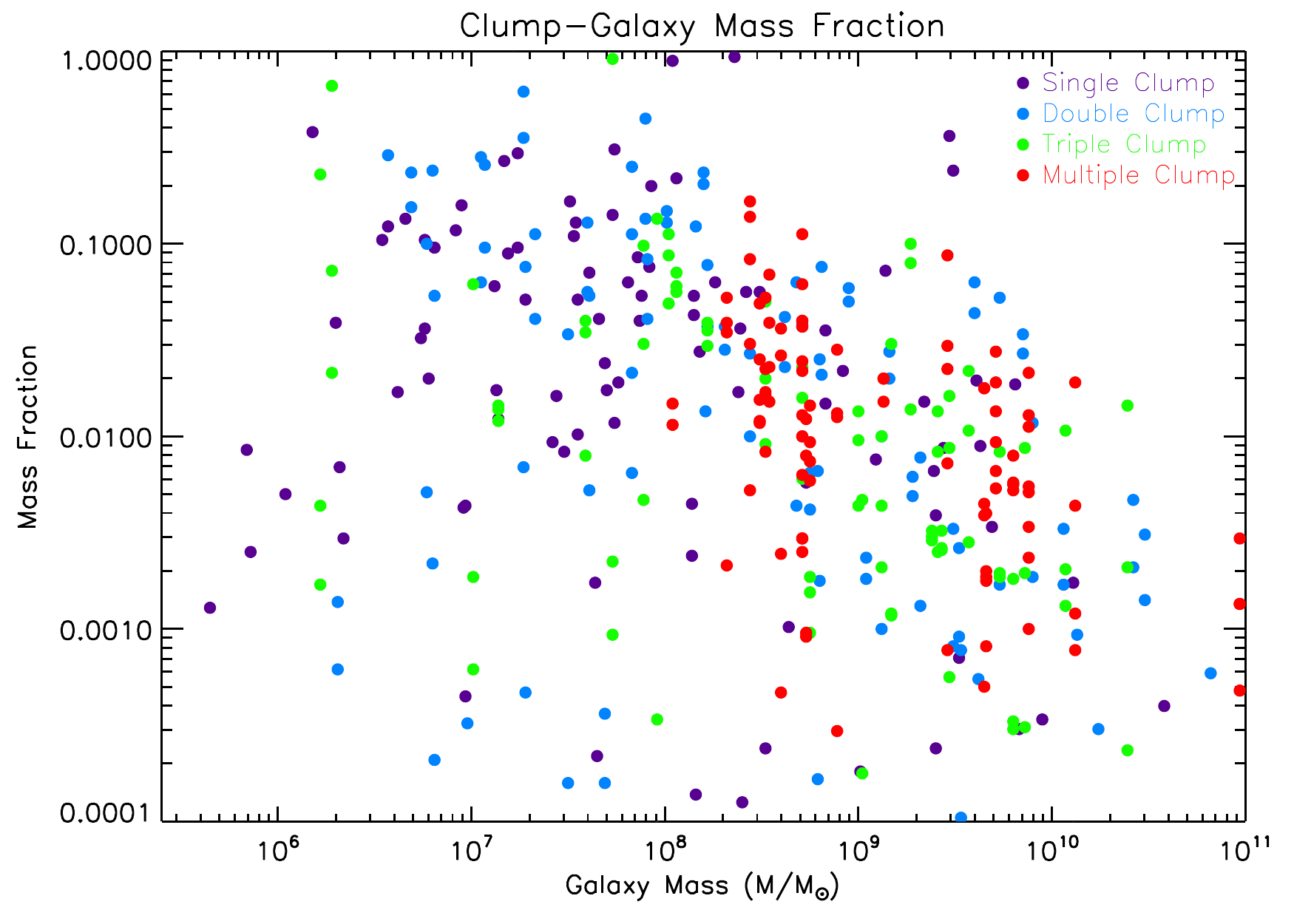}
\caption{Clump mass fraction vs galaxy mass.  The values shown in the plot are the mass fractions for individual clumps and may only comprise a small percentage of the total mass contributed by the clumps.  The median clump contribution to the mass is about 1.2\% and on average about 4.4\% of the total mass is attributed to each individual clump.
\label{massratio}}
\end{figure}

Figure~\ref{massratio} shows our clump-galaxy mass fractions for individual clumps.  Clumps contribute an average of $4.4_{-0.4}^{+0.3}$\% each to the total mass of the galaxy with a median contribution of about $1.2\pm0.2$\%.  This may be insignificant if we only regard single clump systems; however, when we take into account double, triple, and multiple clump systems, which constitute $\sim$50\% of all host galaxies, this contribution from clumps becomes more significant.  This could account for up to 35.2\% of the host galaxy mass in systems with multiple clumps \citep{Wisnioski2012}.  The data also indicates that clumps may contribute more mass in lower mass galaxies as shown by the decreasing trend in mass fraction as galaxy mass increases in Figure~\ref{massratio}.  From simulations, \citet{Mandelker2014} found that each clump contains an average of 1\% of the disk mass, with masses as low as $\sim$0.1\% of the mass of their host disk, and 6\% of the disk SFR, where inner clumps are somewhat more massive and older.  Our results agree remarkably well with their findings for in-situ clumps originating from violent disk instability (VDI).  The mass fractions determined here reach the limits that observations and simulations require to hold clump migration valid as an evolutionary process.

\subsection{Gradients w.r.t. Galactocentric Distance} \label{sec:gradients}
\subsubsection{Mass and Redshift Gradients} \label{sec:smrgrad}

\begin{figure}[ht!]
\epsscale{1.2}
\plotone{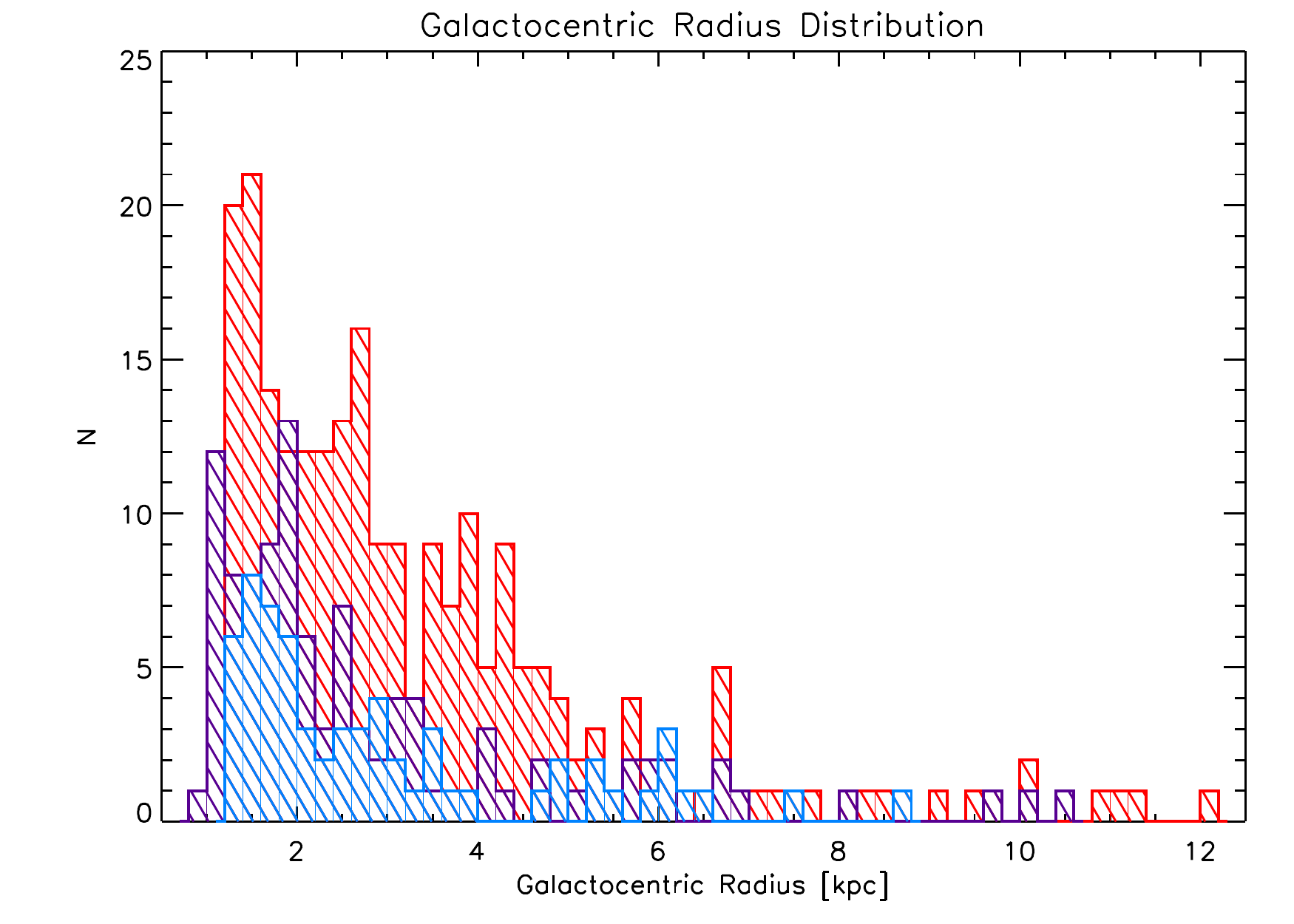}
\caption{Galactocentric radius histograms of clumps based on redshift.  Plots are color-coded based on redshift bin: purple $(0.5<z<0.75)$, blue $(0.75<z<1.0)$, and red $(1.0<z<1.5)$.  
\label{disthist}}
\end{figure}

Correlations between clump mass, age, and SFR with the radial distance to the geometric center of the galaxy, or the galactocentric radius, may also be indicators of clump migration.  Figure~\ref{disthist} shows the distribution of the clumps based on the galactocentric radius depending on redshift.  We find more clumps in the outskirts of galaxies, up to 5 kpc, for galaxies at higher redshifts which is about twice the distance of lower redshift clumps.  This population of high redshift clumps ($z>1.0$) ranging from 3.5 to 5.0 kpc in galactocentric radius is strongly distinguishable in comparison to the distribution of the lower redshift bins.

\begin{figure}[ht!]
\epsscale{1.15}
\centering
\plotone{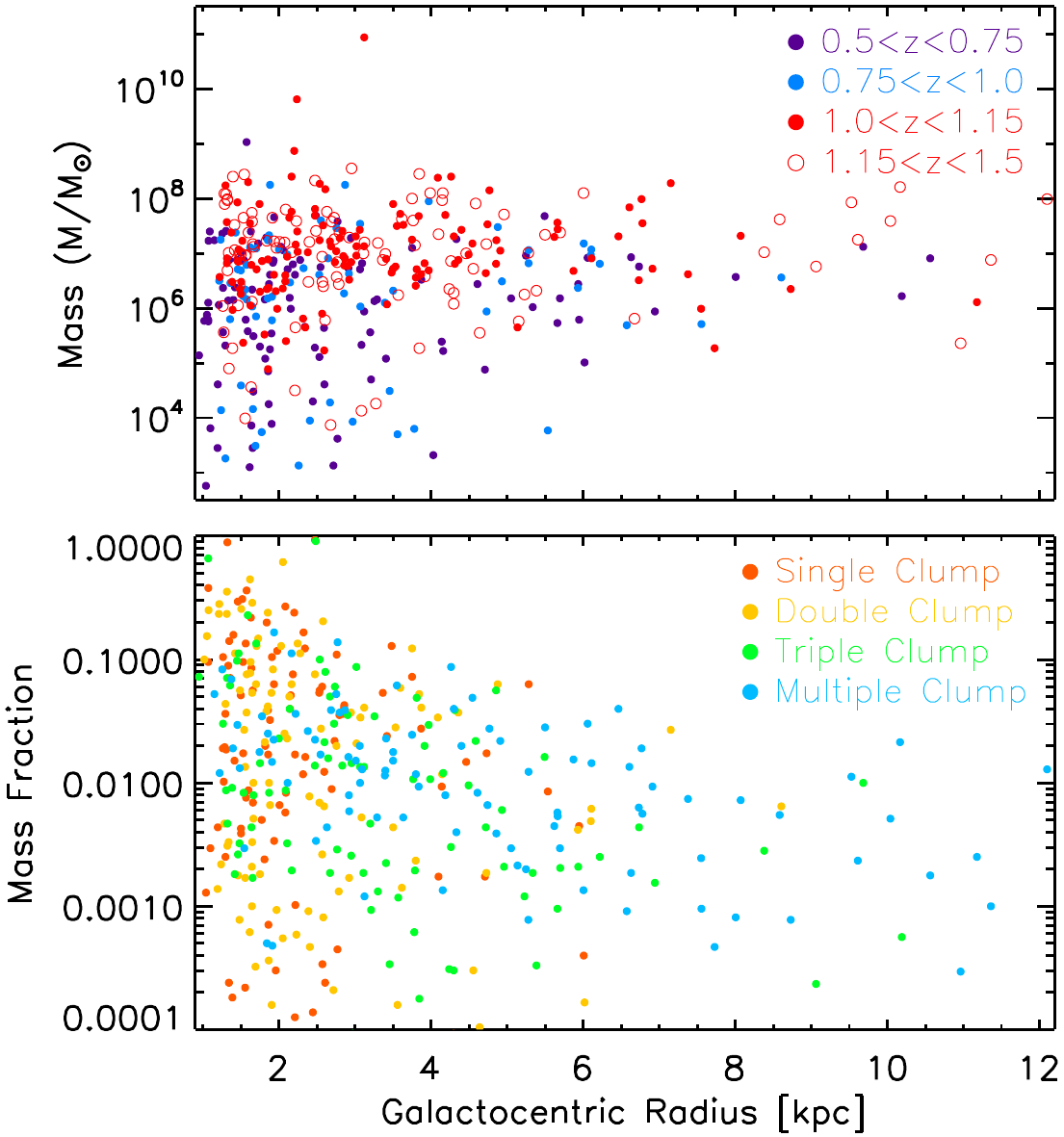}
\caption{(Top) Clump mass vs galactocentric radius.  (Bottom) Mass fraction vs galactocentric radius.  Clump mass fraction increases for clumps located at the inner regions of the host galaxy, with larger contribution from clumps located within a 3 kpc radius.  Clump contribution to the total mass of the galaxy does not exceed 4\% beyond the 6 kpc radii.
\label{massratiodist}}
\end{figure}

Simulations from \citet{Mandelker2014} found that the median mass of in-situ clumps increases the closer the clump is to the center of the galaxy.  The clump mass vs galactocentric radius plot (Fig.~\ref{massratiodist}) does not show this particular trend which agrees with the findings of \citet{Forster2011-sep2}.  Rather we find that lower redshift clumps occupy the lower mass end and high redshift clumps occupy the higher mass end regardless of radius.  We do find that the individual clump mass fraction increases with decreasing galactocentric radius (Fig.~\ref{massratiodist}).  About 94\% of the clumps have a galactocentric radius less than 7 kpc.  When the mass fraction is divided into 2 kpc bins, we find that the median of the mass fraction increases from $0.4^{+0.1}_{-0.2}$\% at a radius of 5-7 kpc up to $2.3^{+0.7}_{-0.3}$\% at a radius of 1-2 kpc.  This is a difference of 5-6 times the median mass budget of the host galaxy occupied by individual clumps in these mass bins.  Although individual clump masses do not follow the trend stated in \citet{Mandelker2014}, clump mass ratios do provide comparable insight into the mass distribution as a function of galactocentric radius.

\subsubsection{Age and SFR Gradients} \label{sec:asfrgrad}

\begin{figure}[ht!]
\epsscale{1.1}
\centering
\plotone{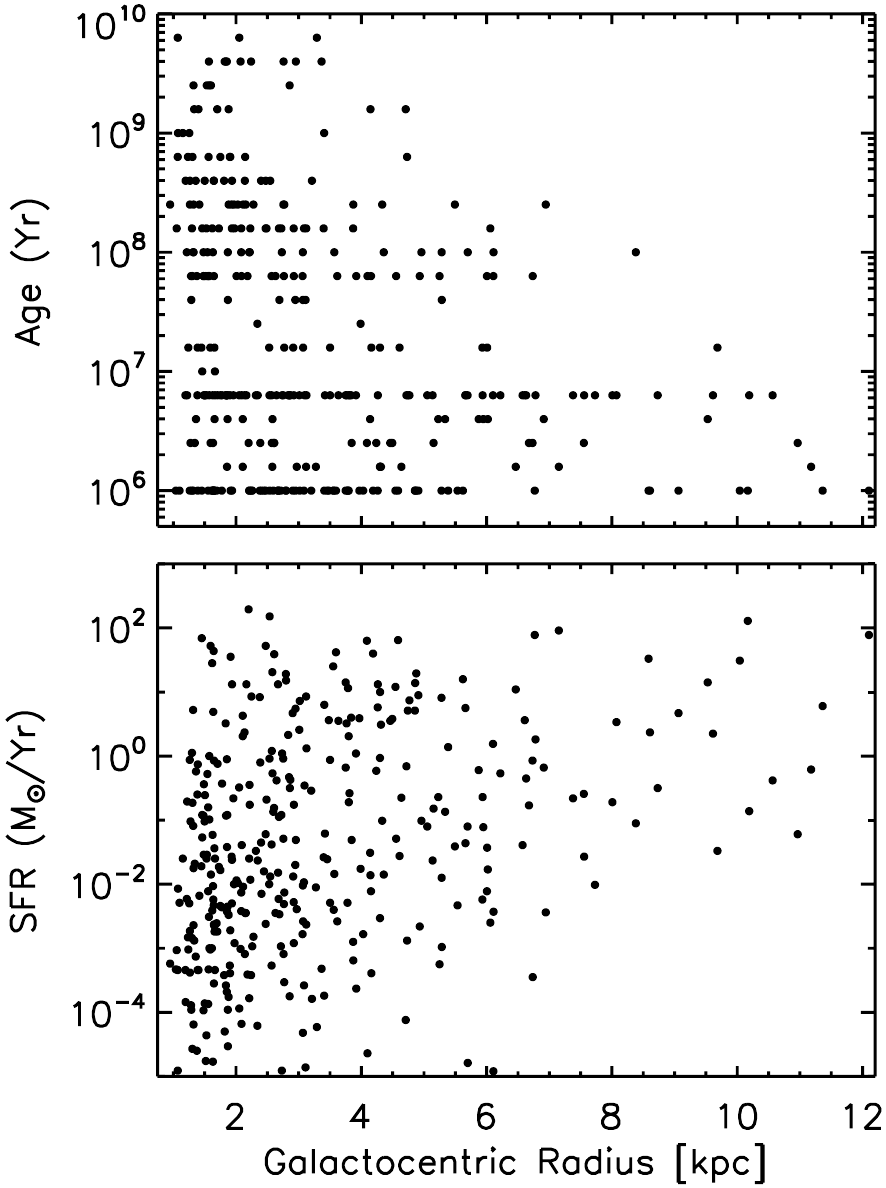}
\caption{(Top) Age vs galactocentric radius of clumps.  Clumps in the young inner region population have an average age of 3.5 Myr while those in the older clump population have an average age of 660Myr. (Bottom) SFR vs galactocentric radius of clumps.  The young clump population in the outer regions of galaxies have greater SFRs (2.24 M$_{\odot}$/yr) with respect to their inner region counterparts (0.59 M$_{\odot}$/yr).
\label{agedistance}}
\end{figure}

We also find a gradient in the clump age vs galactocentric radius (Fig.~\ref{agedistance}) with a younger clump population at greater radii.  We analyze this by dividing the inner region (radius less than 7 kpc) into a young clump population (ages less than 10\textsuperscript{7} yr) and an older clump population (ages greater than 10\textsuperscript{7} yr), and comparing to those in the outskirts (radius greater than 7 kpc).  At radii less than 7 kpc, ages cover a broad range from millions of years to gigayears; however, at radii greater than 7 kpc there are almost no clumps with ages greater than 10 Myr (Fig.~\ref{agedistance}).  The older stellar population at smaller radii could indicate a migration pattern beginning from the outskirts of the galaxy inward toward the center.  Our data supports the findings of observational studies \citep{Forster2011-sep2, Guo2012, Adamo2013} and simulations \citep{Genel2012-jan} where clumps with older stellar populations are closer to the galaxy center.  \citet{Mandelker2014} also find increasing age as galactocentric radius decreases, stating that such results are consistent with the clumps starting to form stars in the outer disk and then gradually migrating inwards.  However, \citet{Genel2012-jan} stated that the observed trend in age gradient, where more distant clumps tend to be younger \citep{Genzel2008, Elmegreen2009b}, is not necessarily an indication for clump migration to the center of the galaxy since ``background'' stars in the clumps could be effecting the ages.  Therefore, we must look for other indicators as to the processes that occur during clump evolution.

The SFR vs galactocentric radius plot (Fig.~\ref{agedistance}) shows a trend of comparatively higher SFR at greater radii, where the young clump population in the outer regions of the host galaxies has a greater SFR (2.24 M$_{\odot}/$yr) with respect to their inner region counterpart (0.59 M$_{\odot}/$yr).  This correlation between SFR and age as a function of galactocentric radius does lead us to infer that highly star-forming regions with young stars of high mass, likely O \& B stars, are forming in the outer regions of galaxies and could be indicative of first generation clumps.  The trends discussed prior, where lower SFR, higher clump mass fractions, and older clumps are found at lower radii, supports the theory that clump migration toward the bulge is occurring.  It is possible to explain the young clumps at lower radii if we assume that they are a second generation population.  Since these would be bright O \& B stars, the UV emission would dominate.  We find that this population has much greater median SFR (0.59 M$_{\odot}/$yr) than the older clump population at lower radii (0.0024 M$_{\odot}/$yr), with a median mass for the young population that is about half the median mass of the older population at lower radii.

\subsection{Rest-frame UV Flux Ratio and Luminosity} \label{sec:fluxlum}

The UV flux ratio is determined by adding the background subtracted flux of all the clumps in the galaxy in the detection band, rest-frame 1500{\AA} flux, and dividing this by the total flux for the galaxy from the isophotal \textit{B}-band flux of the detection band.  We find that each clump typically contributes $\sim$ 5\% of the UV flux with an average of 10\%.  On a larger scale, all clumps contribute $\sim$ 14\% to the host galaxy flux \citep{Adamo2013} with an average contribution of 19\%.  \citet{Elmegreen2005-nov, Elmegreen2009a} derive typically $\sim2\%$ per clump and a total of 25$\%$ on average from rest-frame UV in clump clusters and chains.  They also find that rest-frame UV clumps in more regular galaxies at similar redshifts tend to have lower fractional contributions \citep{Elmegreen2009a}.  \citet{Wuyts2012} required a total UV (rest-frame 2800{\AA}) contribution of 5\% from all clumps at $0.5<z<2.5$ to be considered a clumpy galaxy, which is in agreement with the findings in this study.  \citet{Guo2012} found that individual clumps contribute from 1\%-10\% to the \textit{U}-band and \textit{V}-band, with a median of 5\%, and a total contribution of about 20\% for clumps at $1.5<z<2.5$.  The results discussed here coincide very well with higher redshift studies.

Our results show a higher individual contribution from clumps, twice that found in clumps clusters and chains according to the Elmegreen et al. studies mentioned above.  However, the overall total contribution of the clumps to their host galaxies appears to be somewhat lower.  This could be attributed to the number of clumps typically found in host galaxies in each of the samples due to the different redshifts and the ability to resolve high redshift UV clumps.  The galaxies discussed in \citet{EE2005} for example had redshifts $z\geq1.6$ and contained an average of 10 clumps per galaxy.  Our study of 403 clumps found $\sim2$ clumps per host galaxy (excluding the central bulge) whereas clump clusters and chains discussed in \citet{Elmegreen2009a} usually contained 5-10 clumps.  The difference in the number of clumps per host galaxy explains why although the clumps in our sample individually contribute more of the rest-frame 1500{\AA} flux, the total contributions do not exceed those determined from high-redshift studies.

\begin{figure}[ht!]
\epsscale{1.2}
\plotone{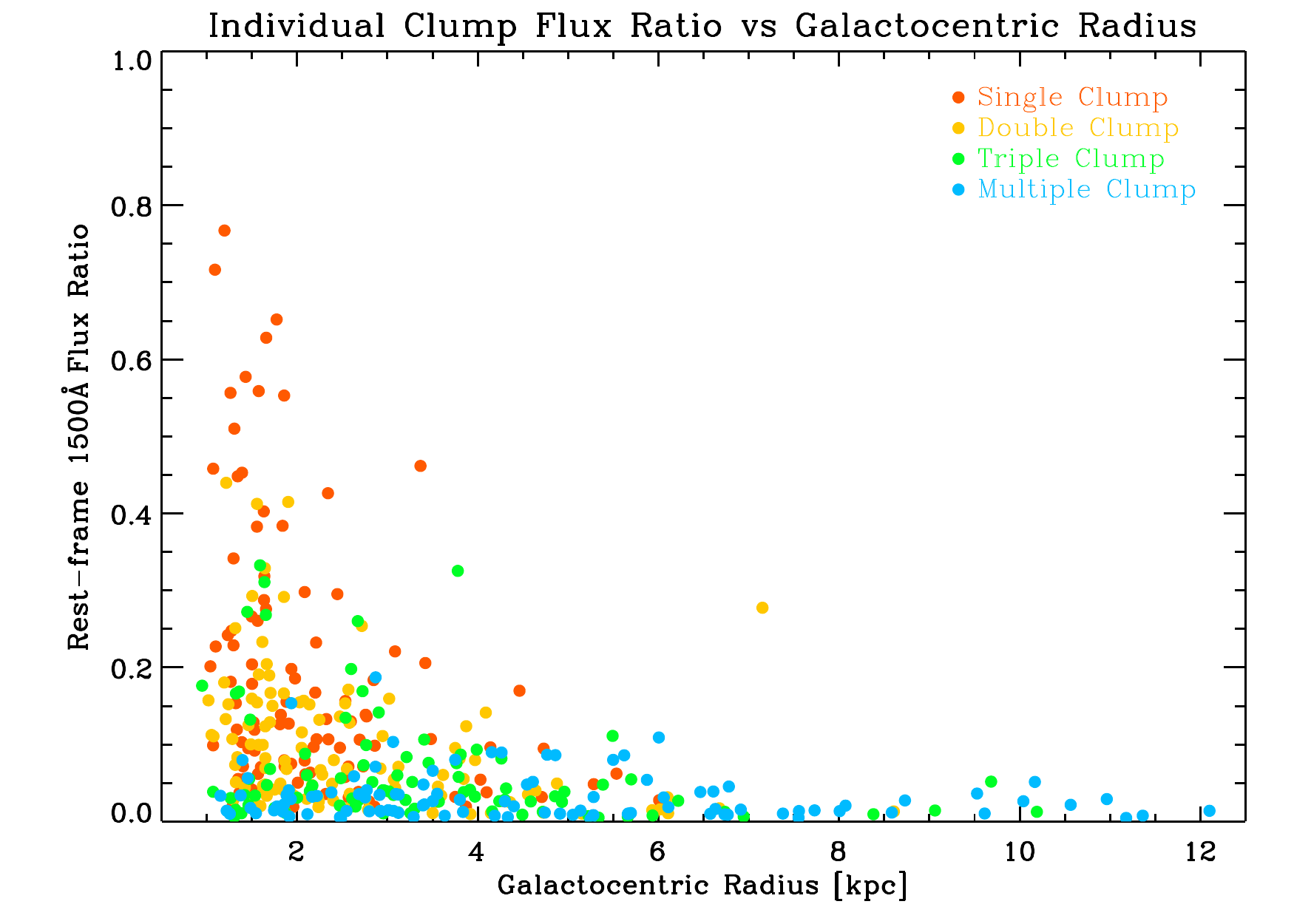}
\caption{Individual clump flux ratio vs galactocentric radius.  Points are color-coded with respect to the number of clumps present in the host galaxy: single clump host galaxies are orange, double clump host galaxies are yellow, galaxies with 3 clumps are green, and galaxies with 4 or more clumps are labeled in blue.  Individual clump flux ratios decrease as a function of radius and clump number.
\label{flux_dist}}
\end{figure}

\begin{figure}[ht!]
\epsscale{1.15}
\plotone{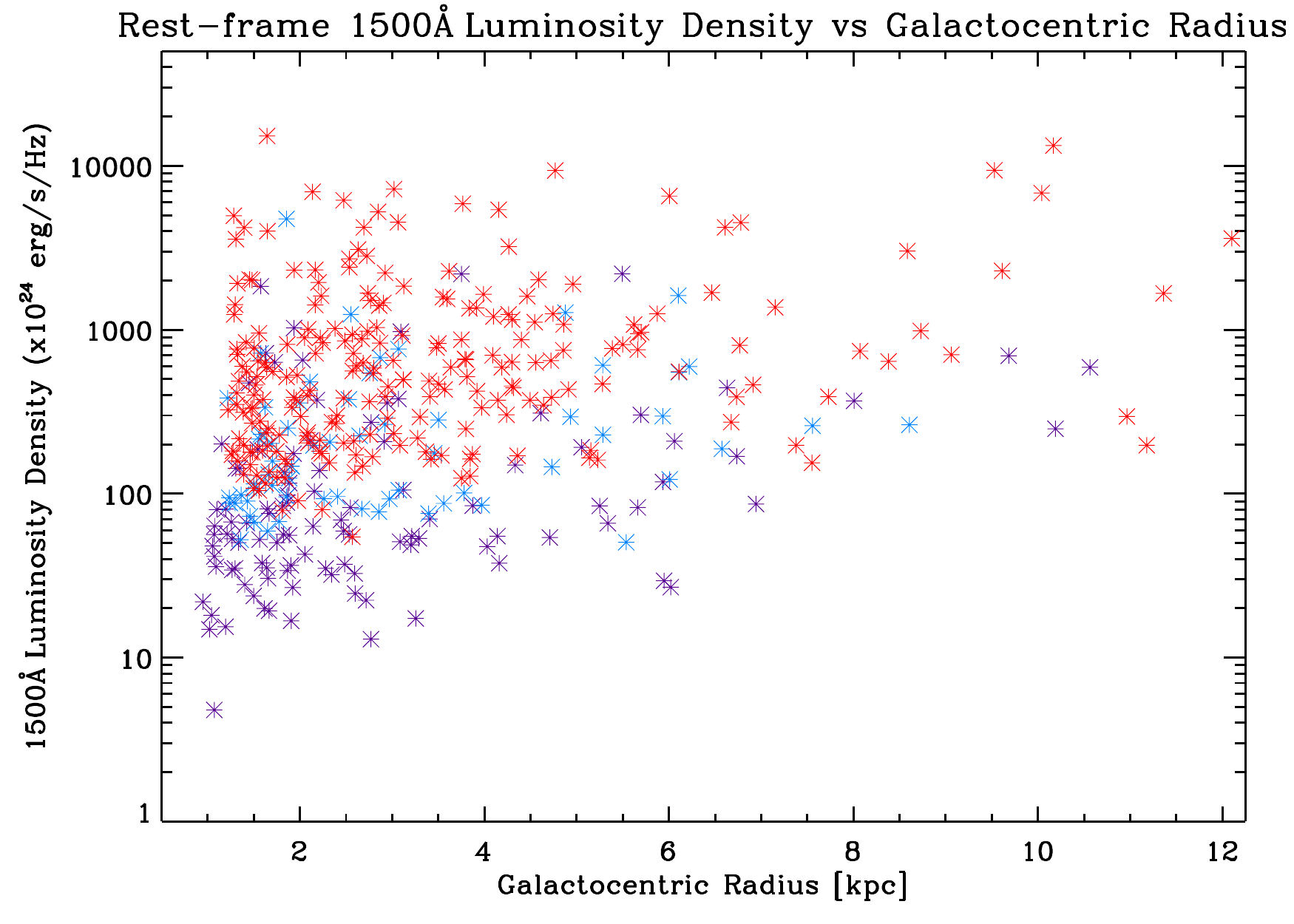}
\caption{Rest-frame 1500{\AA} clump luminosity density vs galactocentric radius.  Luminosity densities determined from background subtracted fluxes for the complete clump sample (403 clumps) are plotted as a function of radius and are color-coded based on the redshift bin scheme of Fig.~\ref{sizemass}.
\label{magdistribution}}
\end{figure}

Our results show a relationship between clump number, flux ratio, and galactocentric radius.  Galaxies with 2 or less clumps, have clumps which individually make up a higher fraction of the rest-frame 1500{\AA} flux and have lower galactocentric radii in comparison to galaxies with 3 or more clumps (Figure~\ref{flux_dist}).  Clumps in galaxies with 3 or more clump detections tend to contribute a smaller fraction of the rest-frame FUV flux and extend out to larger galactocentric radii.  The clumps have rest-frame 1500{\AA} luminosity densities that range from $10\textsuperscript{25}$-$10\textsuperscript{28}$ erg/s/Hz, increasing with respect to redshift (Fig.~\ref{magdistribution}).  The median luminosity density increases from 6.6$\times10\textsuperscript{25}$ erg/s/Hz at $0.5<z<0.75$ to 5.6$\times10\textsuperscript{26}$ erg/s/Hz at $1.0<z<1.5$.  Figure~\ref{magdistribution} also shows that clumps located in the outskirts of galaxies have somewhat higher luminosities (6.9$\times10\textsuperscript{26}$ erg/s/Hz) in comparison to inner region counterparts (2.7$\times10\textsuperscript{26}$ erg/s/Hz).  

\section{Conclusions} \label{sec:conclusion}

We have used $HST/WFC3$ broadband images of 1,403 galaxies from the UVUDF with observed UV and optical photometry to identify 403 clumps in 209 host galaxies at $0.5 \leq z \leq 1.5$.  Deep high resolution WFC3 FUV data allows us to detect and measure kpc-scale clumps through the use of a semi-automated clump finding algorithm.  We measure the physical properties of all clumps through SED-fitting with FAST using seven-band photometry (F225W, F275W, U, B, V, $i$, $z$).  The properties determined from the SED are for a combination of fitting parameters: with exponentially declining SFH, delayed exponentially declining SFH, constrained solar metallicity, and when the metallicity is allowed to float.  The main results of the paper are summarized as follows:

\begin{enumerate}
\item The number of UV-selected clumps in host galaxies varies from predominantly single clump objects to galaxies with up to 8 clumps, with an average number of 2 clumps per galaxy.

\item The host galaxy sample exhibits the linear increasing SFR-stellar mass relation as described in previous studies, indicating that they are typical star-forming galaxies.  The clump SFR-stellar mass relation also exhibits a similar trend.

\item Clumps contribute an average of $\sim$19\% UV flux to the host galaxy, which is a substantial percentage of its UV budget.  Individually, each clump may contribute 5-10\% of the UV flux, which is at least twice that determined from previous studies at higher redshifts.  UV flux ratio also decreases as a function of clump number and galactocentric radius.  If a larger fraction of the UV flux is found in clumps this implies that the impact of clumps in the rest-frame UV morphology are significant but may not be as apparent at other wavelengths.

\item Although clumps contribute a significant fraction of the UV flux budget, individual clumps contribute an average of about 4\% of the total galaxy mass.  The majority of the clumps contribute a combined mass fraction of less than 1\% up to about 40\% of the host galaxy mass.  The UV bright clumps do not dominate the mass budget of the host galaxy.  Thus the main contributors to the mass of the galaxy must be the larger older stellar populations.

\item We find that our results agree remarkably well with those for in-situ clumps formed by violent disk instabilities as described in previous simulation studies and theoretical work, and are also broadly consistent with previous observational studies.

\item The size, mass, redshift, age, and SFR gradients show consistent support of clump migration toward the center of the galaxy.  These properties as a whole allow us to infer the life of clumps.  The distribution of clumps at greater galactocentric radii for high redshift indicate that gravitational instabilities may cause clumps to migrate inward over time.  We find that low redshift clumps are smaller and closer to the galaxy center and higher redshift clumps are typically larger and dominate the population of clumps found in the outskirts of host galaxies.  Additionally, individual clump mass fractions increase with decreasing galactocentric radius, with higher mass fractions at radii less than 3 kpc as compared to clumps at higher radii.  Study of the age gradient led to the analysis of two age populations at lower galactocentric radii and a young population at higher galactocentric radii (greater than 7 kpc) where there are almost no clumps older than 10\textsuperscript{7} years.  We also found a trend toward higher SFR at greater radii, where the young clump population in the outer regions of the host galaxies has a greater SFR with respect to their inner region counterpart.
\end{enumerate}

While some simulations \citep{Tamburello2015} may imply little importance in the role clumps play in galaxy evolution, observational studies such as this work imply otherwise.  Our results, although applicable to UV selected clumps only, are significant.  We show consistent support of clump migration toward the center of the galaxy, but find no strong evidence supporting the quick disruption scenario.  With such a large clump sample (403), we have robust statistical characterization of properties such as clump size, radius, and UV flux ratio.  A larger sample of clumpy galaxies would improve statistics per redshift bin.  Deep IR data with resolution at least comparable to the UVIS data, which does not currently exist, would also be ideal for future study and would consequently provide improved constrained SEDs at greater wavelengths allowing for a more robust comparison of clump properties such as age and SFR.  Further UV studies with HST would be beneficial to future observations with the James Webb Space Telescope (JWST).  Exploring and determining clump properties at $z<2$ would provide important insight into the evolutionary stages of galaxies which could be extended to very high redshift studies as will be conducted with JWST.  However, it is vital that such UV studies be conducted now due to Hubble's limited lifetime.

\acknowledgements

We would like to thank B. Elmegreen for making the UDF clump data from \citet{Elmegreen2013} available for Figure~\ref{sizemass}.  We also thank K. Whitaker for her assistance in mastering FAST, N. Bond, Y. Guo, and C. Leitherer for valuable discussions, and A. Fitzmaurice for her contributions to the earlier stages of this work.  This material is based upon work supported by the National Aeronautics and Space Administration under Grant Number NNX13AT09H issued through the NASA Education Minority University Research Education Project (MUREP) through the NASA Harriett G. Jenkins Graduate Fellowship activity.  DFdM was supported by STScI grant number HST-GO-12534.008-A.

\clearpage

\appendix
\section{UV Resolution vs PSF-Matched Resolution Comparison} \label{sec:appendix}

\begin{figure}[ht!]
\centering
\includegraphics[width=8.5cm]{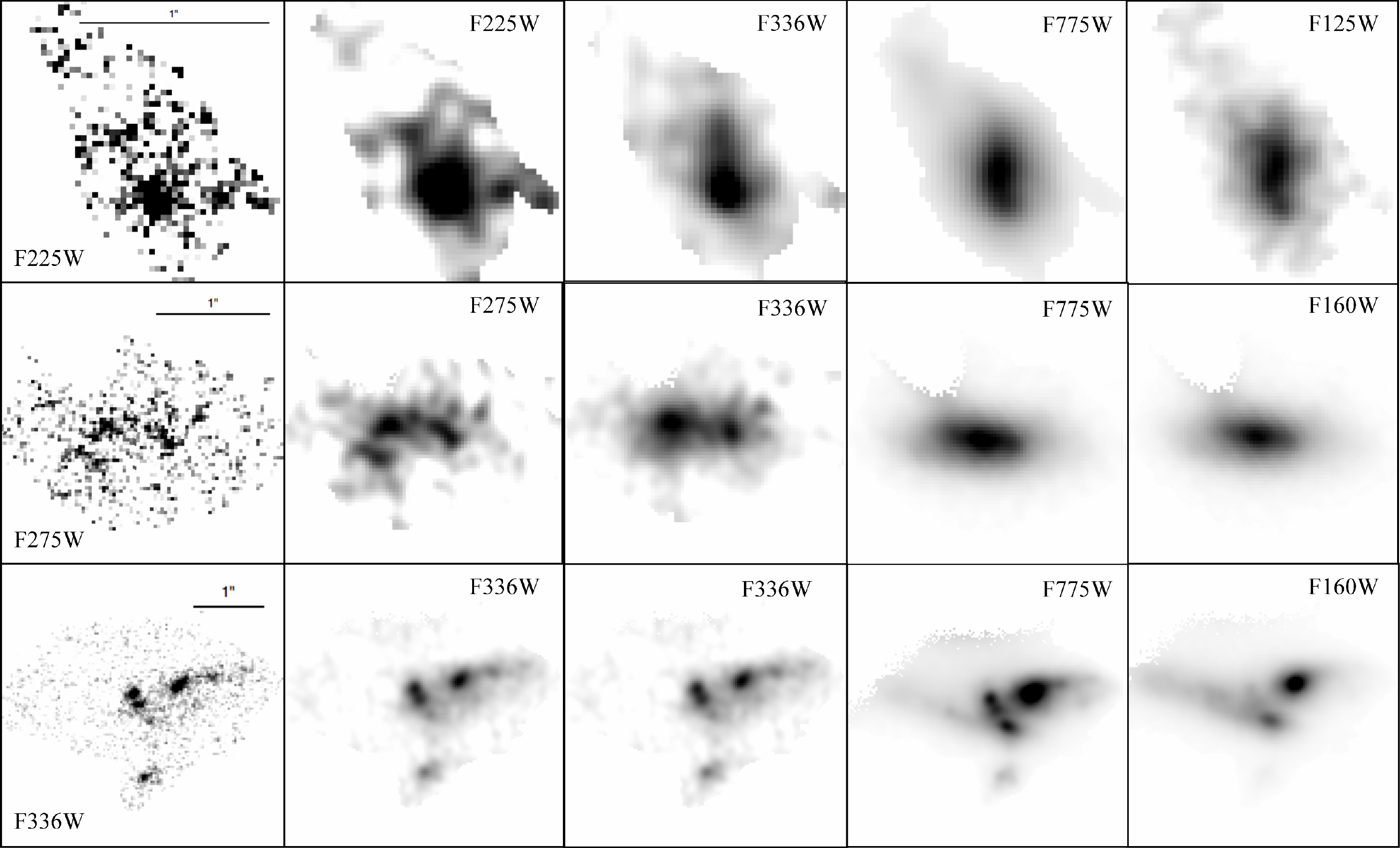}
\caption{UV and Optical HST Resolution vs H-band PSF-matched Mosaics. Original resolution detection band images are shown in the left column with the psf-matched images of the same filter shown directly to the right.  The two middle columns show the observed $U$-band and $i$-band psf-matched images.  The last column provides the psf-matched image of the highest IR data available.  Redshifts from top to bottom are $z =$ 0.747, 0.972, and 1.047.
\label{psfgallery}}
\end{figure}

One of the challenges in this study is addressing the validity of our derived physical properties for the clumps when using only the 3 UV and 4 optical observed passbands (without the NIR data) for clump photometry.  The morphology of galaxies is observed to change drastically between the UV and IR passbands, often times appearing clumpy and disjointed in the UV bands, and smooth and symmetric in the IR bands \citep{Petty2014, Petty2009}, absent of clearly visible clumps.  We omit the IR passbands because clumps were not observed or unresolved in the NIR.  The image resolution is a key factor for this choice because the IR resolution (FWHM $\sim$0.20$''$) is much lower than the UV and optical HST resolution (FWHM $\sim$0.10$''$).  We validate our derived physical properties from SED fitting by comparing the resultant physical parameters obtained with either UV through optical or UV through NIR photometry for the clump sample.  With these test cases, or rather test-clumps, we compare the clump properties at the UV and optical HST (UVO HST) resolution data to psf-matched UV+optical (UVO) and psf-matched UV+optical+IR (UVOIR) resolution images that include F105W, F125W, F140W, and/or F160W when available (Figure~\ref{psfgallery}), with all bands psf-matched to F160W when the NIR is used.  This comparison is performed for the same two metallicity cases described in the paper using an exponentially declining SFH.

\begin{figure}[ht!]
\epsscale{1.15}
\plottwo{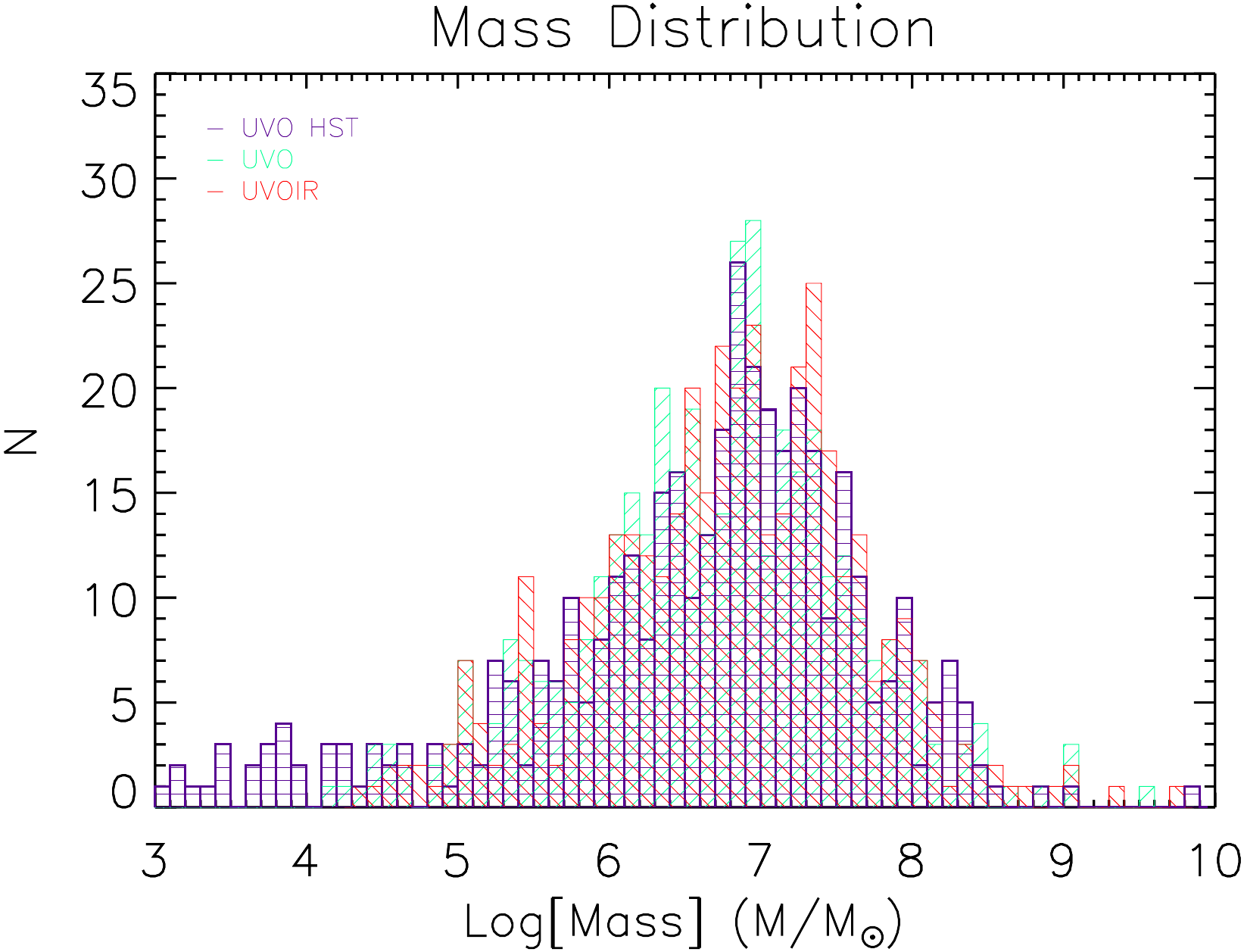}{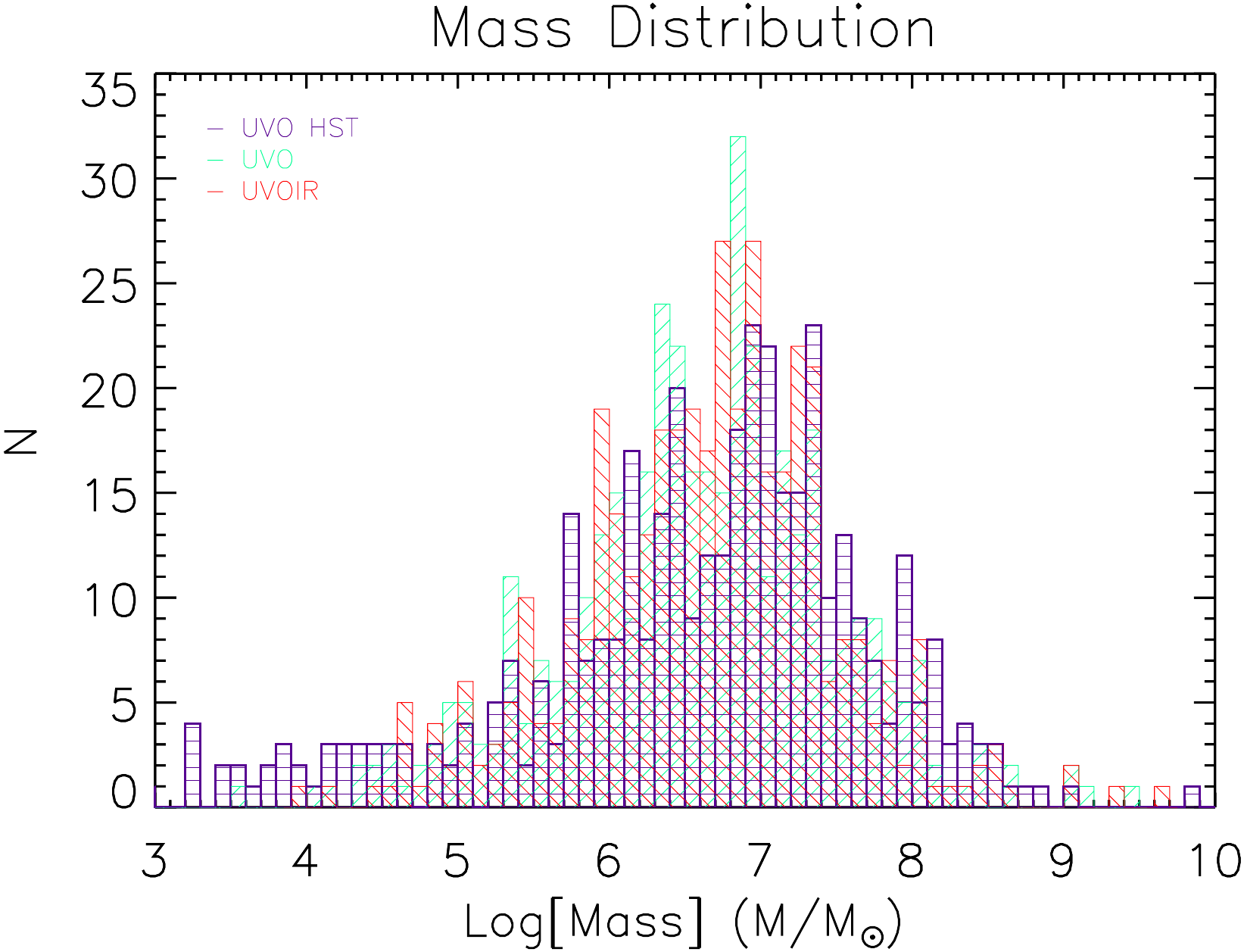}
\caption{The mass distribution (modeled using EXP SFH with FAST) of all the test clumps at UV resolution (purple) with psf-matched resolution UV+optical (green) and psf-matched resolution UV+optical+IR (red) included for a constrained solar metallicity (left) and when the metallicity is allowed to float (right).  The bin size is 0.10.  The mass distribution peaks at about 8-10 million M$_{\odot}$ for both metallicity models.\\
- Test clumps are the UV detected clumps for an exponentially declining SFH which are used for comparison to determine disparities in the data that may arise from using the UV and optical bands only and from using psf-matched images.
\label{testmassdist}}
\end{figure}

\begin{figure}[ht!]
\epsscale{0.8}
\plotone{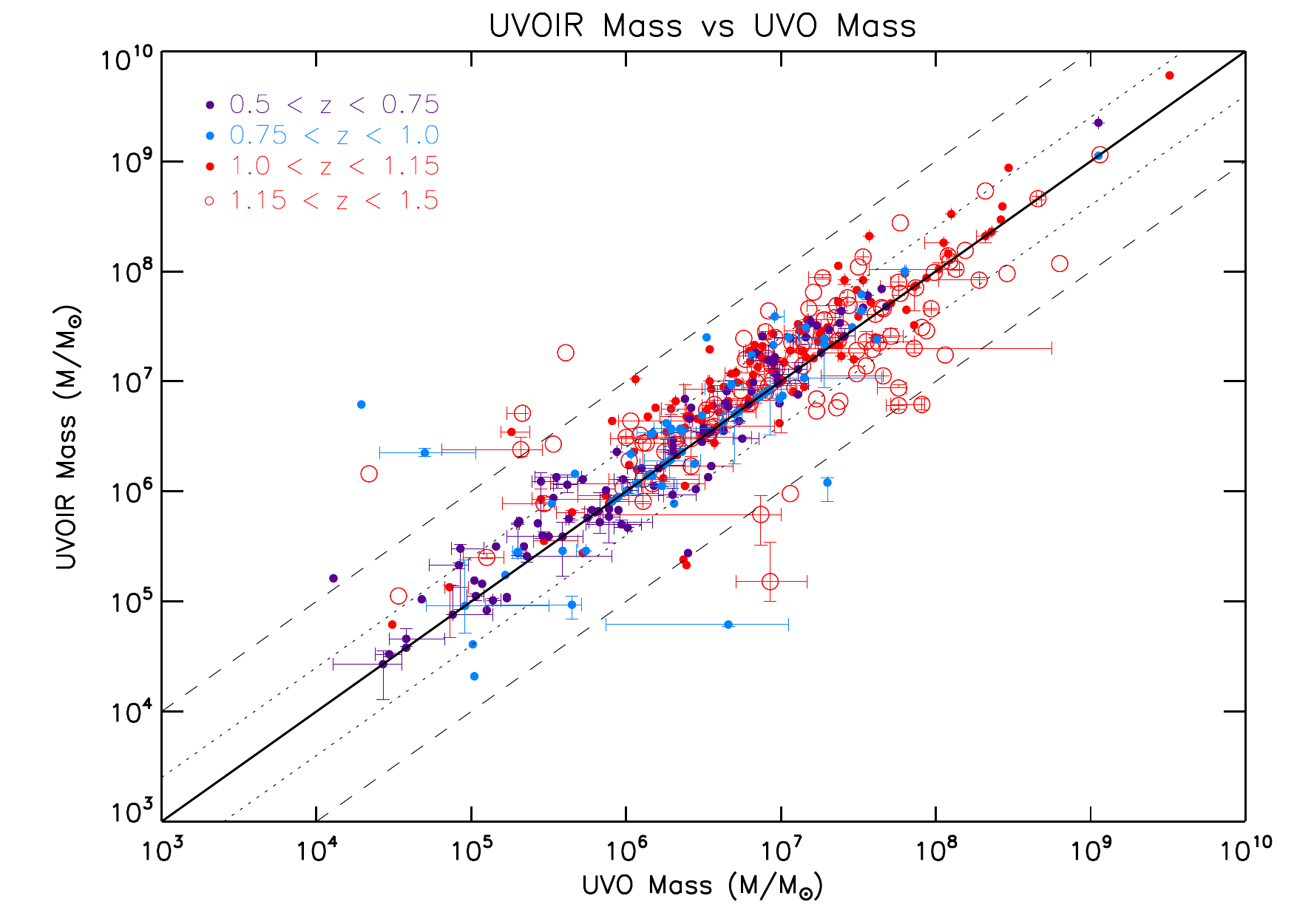}
\caption{Comparison of the masses determined from SED fitting at solar metallicity inclusive and exclusive of IR data from psf-matched images.  Clump mass is color-coded based on redshift: $0.5 < z < 0.75$ in purple, $0.75 < z < 1.0$ in blue, $1.0 < z < 1.15$ in solid red circles, and $1.15 < z < 1.5$ in open red circles.  The solid black line represents the 1:1 ratio that would exist if the masses inclusive and exclusive of the IR were identical.  Error bars are provided for half of the test sample.  There is good correlation regardless of redshift, with 96\% of clump masses being within 1.0 dex (dashed line) of the 1:1 ratio and with median scatter of 0.4 dex (dotted line).  
\label{massmass}}
\end{figure}

Figure~\ref{testmassdist} shows how the masses at the psf-matched resolution compare with the mass results of the same sample at the UVO HST resolution.  The mass distribution peaks at about 8-10 million M$_{\odot}$ for both metallicity cases, with smaller peaks present in both.  When the clump metallicity is constrained to solar metallicity, the peak of the mass distribution for the UVOIR sample does occur at slightly greater mass (about 30 million M$_{\odot}$); however, the complete mass distribution inclusive of the IR does not fully shift toward higher masses.  When the metallicity is allowed to float the histogram does show a small systematic shift in the mass without the NIR data; however, it is not a drastic shift.  Similarly, \citet{Buat2014} compared stellar properties derived using combinations of UV, NIR, and IR data and found that omitting NIR data would lower the mass by an average of 15\% and that omission of IR data would have even less of an impact for galaxies at $z > 1$.  Table~\ref{massdiff} shows that when the NIR is excluded the average and median masses are lower for both metallicity models.  We find differences in the median mass of the psf-matched resolution UVO and UVOIR of about 15\% for both metallicity cases (Table~\ref{massdiff}) and correlation coefficients of 0.96 for solar metallicity and 0.97 for floating metallicity.  The median mass difference agrees very well with the conclusions from \citet{Buat2014}; however, as shown in Figure~\ref{massmass}, clumps at $z > 1$ are impacted just as equally as those at lower redshift.  Figure~\ref{massmass} does exhibit a very good correlation regardless of redshift, with 96\% of the clump masses being within 1.0 dex of the 1:1 linear ratio and having a median scatter of 0.4 dex.

\begin{deluxetable*}{c|ccc|ccc}
\tabletypesize{\footnotesize}
\tablewidth{0pt}
\tablecolumns{7}
\tablenum{5}
\tablecaption{Mass Comparison Table \label{massdiff}}
\tablehead{\colhead{Metallicity} & \colhead{} & \colhead{Median ($10^{6}$ M$_{\odot}$)} & \colhead{} & \colhead{} & \colhead{Average ($10^{6}$ M$_{\odot}$)} & \colhead{}}
\hline
& UVOIR & UVO  & UVO & UVOIR  & UVO & UVO$^{\dagger}$ \\
& (PSF-matched) & (PSF-matched) & (Original) & (PSF-matched) & (PSF-matched) & (Original) \\
\hline
&&&&&&\\
Solar & 6.76$^{+1.00}_{-0.59}$ & 5.75$^{+0.70}_{-0.50}$ & 6.46$^{+0.62}_{-0.83}$ & 53.14$^{+12.7}_{-16.0}$ & 38.57$^{+7.44}_{-9.53}$ & 261$^{+138}_{-212}$ \\
&&&&&&\\
Float & 5.25$^{+0.51}_{-0.68}$ & 4.47$^{+0.90}_{-0.66}$ & 6.31$^{+0.93}_{-0.94}$ & 40.34$^{+10.45}_{-9.95}$ & 35.18$^{+7.74}_{-7.66}$ & 264$^{+137}_{-211}$ \\
\tablecomments{UVOIR is UV+optical+IR images and UVO is UV+optical images.  Original denotes the results obtained using only original resolution images and PSF-matched denotes the results obtained using psf-matched degraded resolution images.\\
                           Metallicity parameters for data presented are as follows: Z$_{\odot}$ (solar metallicity) and Z$_{Float}$=[0.20Z$_{\odot}$, 0.40Z$_{\odot}$, Z$_{\odot}$] (floating metallicity). \\
                           ${\dagger}$When clumps with fluxes in less than 7 filters are excluded the averages and uncertainties are $33.69^{+4.84}_{-4.84} \times10^{6}$ M$_{\odot}$ and $40.32^{+4.62}_{-4.84} \times 10^{6}$ M$_{\odot}$ for solar and floating metallicity respectively.}
\end{deluxetable*}

\begin{figure}[ht!]
\epsscale{0.8}
\plotone{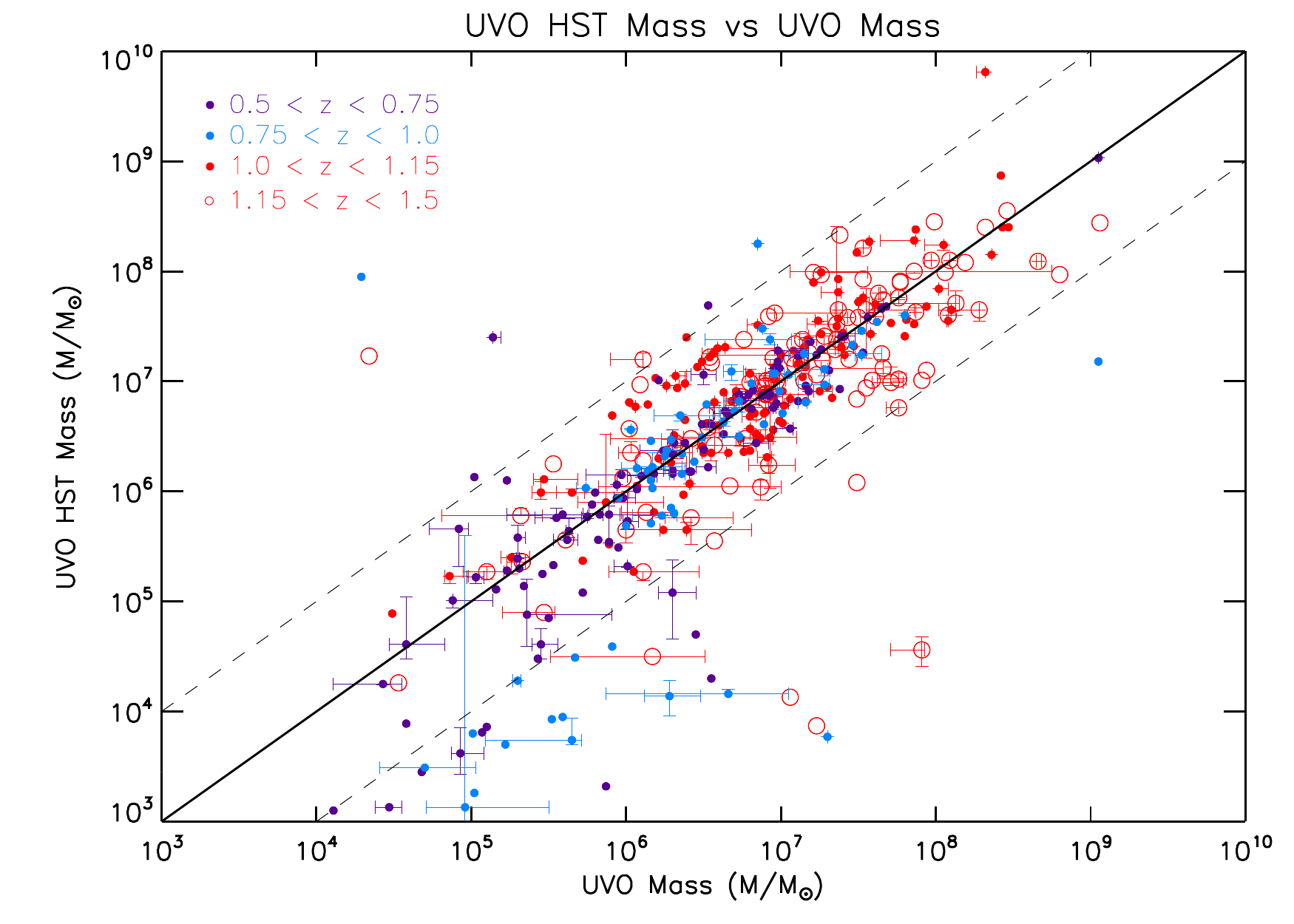}
\caption{Comparison of the masses determined with UV+optical data at the original WFC3 resolution and when psf-matched to the H-band for solar metallicity.  Clump mass is color-coded based on the redshift scheme detailed in Figure~\ref{massmass}.  The solid black line represents the 1:1 ratio between the two mass results which should be present if no effect from resolution differences arise and the dashed lines represent a 1.0 dex scatter from the 1:1 linear ratio.  Error bars are provided for half of the test sample.  Most of the test clump masses fall within the spread; however, there are some outliers, mainly below the bottom dashed line.
\label{massres}}
\end{figure}

\begin{figure}[ht!]
\epsscale{0.8}
\plotone{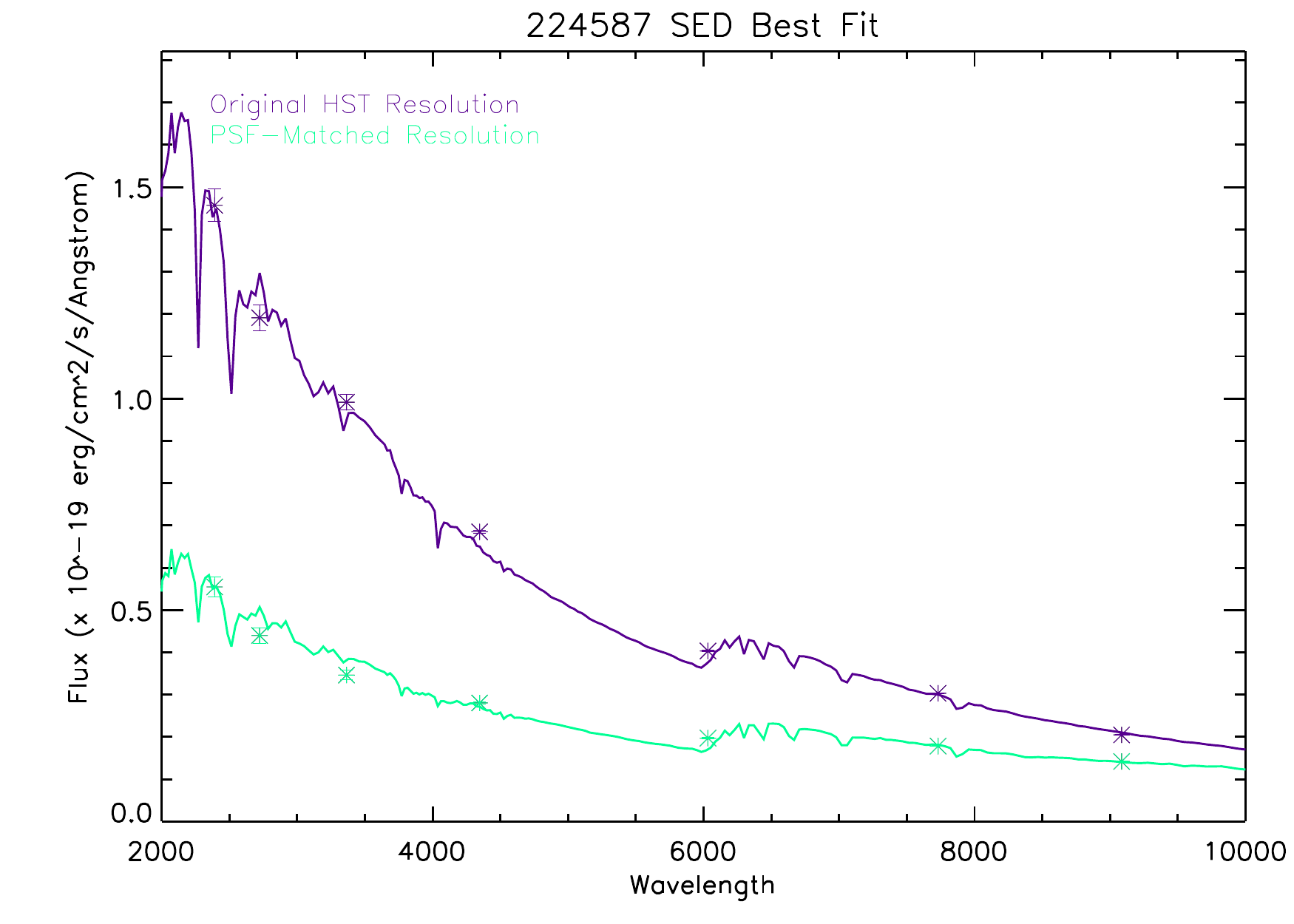}
\caption{Original HST Resolution vs PSF-Matched Resolution SED Fits.  The SED shown here is for galaxy ID 24587 clump \#2, with the fit for original resolution in purple, the psf-matched resolution in green, and the respective data points overplotted.  The ratio between the two resolutions across the SED varies, with a smaller ratio at lower wavelength, near the observed UV.  The mass determined at the original resolution is 3.7 million M$_{\odot}$, while at psf-matched resolution 11.5 million M$_{\odot}$.}
\label{sed}
\end{figure}

We investigate the effects of image resolution on the masses determined from SED fitting.  Figure~\ref{massres} illustrates the effects of utilizing the UVO HST resolution images in comparison to H-band psf-matched resolution images.  Table \ref{massdiff} shows that the averages for the psf-matched and original resolution UVO masses are quite different with rather high uncertainties; however, when limited to the sample of clumps which have photometry in all 7 passbands (69\% of all clumps), the uncertainties are on the order of $\pm 5.0$ (see footnote for Table \ref{massdiff}).  Clumps with less than 7 filters tend to have higher mass estimates, which drives the overall average and uncertainties to higher values.  We find that 90\% of the test clumps are within the 1.0 dex scatter.  There are several outliers beyond this limit, mainly below the 1:1 ratio line, which indicates that the masses determined using the psf-matched images are greater than the masses determined using the original HST resolution photometry.  The scatter is largest for clumps where the UVO HST resolution mass is less than 10\textsuperscript{5} M$_{\odot}$.  Further investigation of the outliers shows that these particular clumps are missing data from 2 or more passbands for the UVO HST SED fit.  Therefore, the SED fit is being determined by 5 or less passbands, whereas the UVO psf-matched images still provide small fluxes for the missing passbands accounting for the higher masses.  The fluxes in the UVO psf-matched images may be present as a result of poor deblending or as a result of de-convolving when psf-matching to lower resolution.  Disparities in the photometry between the original HST resolution and the psf-matched resolution, that account for clumps where the scatter is above the 1:1 linear ratio, may arise as a result of psf-matching.  When psf-matched more of the flux contained in the border/wings of the clump is lost and therefore result in smaller masses when performing SED fitting as illustrated in Figure~\ref{sed}.  The SED fit in Figure~\ref{sed} shows a smaller ratio between fluxes in the observed UV of both resolutions in comparison to the ratio at higher wavelengths leading to the differences in mass observed in Figure~\ref{massres} for the outliers above the linear ratio.  This shows the difficulty in measuring the photometry of small clumps in the lower resolution psf-matched images, and is the reason for omitting the NIR data in this study.

Overall, the masses determined from these tests show that the disparities seem to primarily arise between the UV resolution and F160W psf-matched resolution results.  The information provided in this Appendix shows a clear comparison of the UVO HST resolution data and the H-band psf-matched UVO resolution data, as well as a clear comparison of the H-band psf-matched data when excluding and including the IR.  The difficulty in measuring clump photometry in reduced resolution images combined with the lack of ability to detect clumps in the NIR justifies our leaving out the NIR photometry in our clump SED fitting.  We conclude that while the omission of the NIR data does cause a slight systematic shift in the masses, the resolution effects are much stronger than this and therefore we omit the NIR data.  Future observations at higher resolution, such as with JWST, will enable more precise clump properties.

\clearpage

\bibliographystyle{apj}
\bibliography{research}

\end{document}